\documentclass[prd,aps,a4paper,nofootinbib,showpacs,showkeys]{revtex4}  

\newif\ifusesec
\usesectrue  
   
\usepackage{graphicx} 
\usepackage{mathrsfs}
\usepackage{latexsym,amsmath,amsfonts,amssymb}
\usepackage{multirow}

\newcommand{\beq}{\begin{equation}}
\newcommand{\eeq}{\end{equation}}

\begin{document}

\title{Relativistic Gravity Gradiometry: The Mashhoon--Theiss Effect}

\author{Donato \surname{Bini}$^1$}
\author{Bahram \surname{Mashhoon}$^2$}

\affiliation{$^1$Istituto per le Applicazioni del Calcolo ``M. Picone'', CNR, I-00185 Rome, Italy\\
$^2$Department of Physics and Astronomy,\\
University of Missouri, Columbia, Missouri 65211, USA}
 
\date{\today}

\begin{abstract}
In general relativity, relativistic gravity gradiometry involves the measurement of the relativistic tidal matrix, which is theoretically  obtained from the projection of the Riemann curvature tensor onto the orthonormal tetrad frame of an observer. The observer's 4-velocity vector defines its local temporal axis and its local spatial frame is defined by a set of three orthonormal nonrotating gyro directions. The general tidal matrix for the timelike geodesics of Kerr spacetime has been calculated by Marck~\cite{Marck}.  We are interested in  the measured components of the curvature tensor along the inclined ``circular" geodesic orbit of a test mass about a slowly rotating astronomical  object of mass $M$ and angular momentum $J$. Therefore, we specialize Marck's results to such a ``circular" orbit  that is tilted with respect to the equatorial plane of the Kerr source. To linear order in $J$, we recover the Mashhoon--Theiss effect~\cite{MaTh}, which is due to a small denominator (``resonance") phenomenon involving the frequency of geodetic precession. The Mashhoon--Theiss effect shows up as a special long-period gravitomagnetic part of the relativistic tidal matrix; moreover, the effect's short-term manifestations are contained in certain post-Newtonian secular terms.  The physical interpretation of this gravitomagnetic beat  phenomenon is briefly discussed.  
\end{abstract}

\pacs{04.20.Cv, 04.25.Nx, 04.80.-y}
\keywords{relativistic gravity gradiometry, post-Schwarzschild approximation, Kerr spacetime}
\maketitle

\section{Introduction}

In Newton's theory of gravitation, gravity gradiometry involves the measurement and study of the variations in the acceleration of gravity. Imagine two \emph{nearby} test masses $m_a$ and $m_b$ falling freely in the gravitational potential $U$ of external sources.  Let $\boldsymbol{\xi} = \mathbf{x}_a(t) - \mathbf{x}_b(t)$ be the instantaneous deviation between the trajectories of the two neighboring masses; then, it follows from Newton's second law of motion that to linear order in 
$\boldsymbol{\xi}$, 
\begin{equation}\label{I1}
\frac{d^2\xi^i}{dt^2}+\kappa^i{}_j\, \xi^j=0\,,
\end{equation}
where, in this tidal equation, $\kappa_{ij}$ is the Newtonian tidal matrix,
\begin{equation}\label{I2}
\kappa_{ij}(\mathbf{x})= \frac{\partial^2 U}{\partial x^i\partial x^j}\,,
\end{equation}
evaluated along, say, $\mathbf{x}_a(t)$, taken to be the reference trajectory.  In the source-free region under consideration here, Poisson's equation for $U$ reduces to Laplace's equation $\nabla^2 U=0$. The tidal matrix is thus symmetric and traceless; moreover, each element of the Newtonian tidal matrix is a \emph{harmonic} function, since in this case $\nabla^2 \kappa_{ij}=0$. It is clear from the tidal Eq.~\eqref{I1} that when tides dominate, the shape of a tidally deformed body would generally tend to either a cigar-like or a pancake-like configuration, since the traceless tidal matrix can in general have either two positive and one negative or  two negative and one positive  eigenvalues, respectively. It is important to note that the tidal matrix is completely independent of the test masses as a consequence of the universality of gravitational interaction, namely, the principle of equivalence of gravitational and inertial masses. Historically, the science of gravity gradiometry goes back to E\"otv\"os, who used a torsion-balance method to test the principle of equivalence (1889--1922). The magnitude of a gravity gradient is expressed in units of  E\"otv\"os, 1 E =$10^{-9}$ s$^{-2}$.

Gravity gradiometry has many significant practical applications. Furthermore, gravity gradiometers  of high sensitivity have been developed that are suitable for use in basic physics experiments. In this connection, we must mention the highly sensitive superconducting gravity gradiometer developed by Paik {\it et al.}, which employs superconducting quantum interference devices (SQUIDs)~\cite{HJP1,HJP2,HJP3}.  Regarding possible future space applications, gravity gradiometry has also become possible using atom interferometry~\cite{AT1, AT2}. 

Relativistic gravity gradiometry involves the measurement of the Riemannian curvature of spacetime. In Einstein's general relativity (GR), the gravitational field is represented by the spacetime curvature. When an observer measures a  gravitational field, the curvature tensor must be projected onto the tetrad frame of the observer. Imagine an observer following a  future directed  timelike geodesic world line  $x^\mu(\tau)$ in spacetime, where $\tau$ is the proper time of the observer along the geodesic. The observer carries an orthonormal parallel-propagated tetrad frame $\lambda^\mu{}_{\hat \alpha}$, 
\begin{equation}\label{I3}
g_{\mu \nu} \,\lambda^\mu{}_{\hat \alpha}\,\lambda^\nu{}_{\hat \beta}= \eta_{\hat \alpha \hat \beta}\,,\qquad \frac{D \lambda^\mu{}_{\hat \alpha}}{d\tau}=0\,,
\end{equation}
where $\lambda^\mu{}_{\hat 0}=dx^\mu/d\tau$ is the unit timelike  tangent vector that is the observer's 4-velocity  and $\lambda^\mu{}_{\hat i}$, $i=1,2,3$, are unit spacelike ``gyroscope" directions that form the local spatial frame of the observer. In display~\eqref{I3}, 
$\eta_{\hat \alpha \hat \beta}$ is the Minkowski metric given by diag$(-1, 1, 1, 1)$; in our convention, the signature of the metric is +2 and we choose units  such that $G=c=1$, unless specified otherwise. The measured components of the Riemann tensor are then the scalars given by
\begin{equation}\label{I4}
R_{\hat \alpha \hat \beta \hat \gamma \hat \delta} = R_{\mu \nu \rho \sigma} \,\lambda^\mu{}_{\hat \alpha}\,\lambda^\nu{}_{\hat \beta}\,\lambda^\rho{}_{\hat \gamma}\,\lambda^\sigma{}_{\hat \delta}\,.
\end{equation}

It is interesting to take into account the symmetries of the Riemann tensor and express Eq.~\eqref{I4} in the standard manner as a $6\times 6$ matrix $(R_{AB})$, where $A$ and $B$ are indices that belong to the set $\{01,02,03,23,31,12\}$. The general form of this matrix is 
\beq\label{I5}
\left[
\begin{array}{cc}
\mathcal {E} & \mathcal {H}\cr
\mathcal {H}^{\dagger} & \mathcal {S}\cr 
\end{array}
\right]\,,
\eeq
where $\mathcal {E}$ and $\mathcal {S}$ are symmetric $3\times 3$ matrices and $\mathcal{H}$ is traceless. Here, $\mathcal {H}^{\dagger}$ is the transpose of matrix $\mathcal {H}$. The relativistic tidal matrix $\mathcal{E}$ represents the measured gravitoelectric components of the Riemann curvature tensor, while $\mathcal{H}$ and  $\mathcal{S}$ represent its gravitomagnetic and spatial components, respectively~\cite{Matt, GEM}. In the vacuum region exterior to material sources and free of nongravitational fields, the spacetime is Ricci flat as a consequence of Einstein's field equations of GR and the measured components of the curvature tensor are then given by
\beq\label{I6}
\left[
\begin{array}{cc}
{\mathcal E} & {\mathcal H}\cr
{\mathcal H} & -{\mathcal E}\cr 
\end{array}
\right]\,,
\eeq
where  $ \mathcal{E}$ and $\mathcal {H}$ are symmetric and traceless. 
In this case, the Riemann curvature tensor degenerates into the Weyl conformal curvature tensor whose gravitoelectric and gravitomagnetic components are then
\begin{equation}\label{I7a} 
\mathcal{E}_{\hat a \hat b}= C_{\alpha\beta\gamma\delta}\,\lambda^\alpha{}_{\hat 0}\,\lambda^\beta{}_{\hat a}\, \lambda^\gamma{}_{\hat 0}\, \lambda^\delta{}_{\hat b}\,,\qquad \mathcal{H}_{\hat a \hat b}=C^*_{\alpha\beta\gamma\delta}\,\lambda^\alpha{}_{\hat 0}\,\lambda^\beta{}_{\hat a}\, \lambda^\gamma{}_{\hat 0}\, \lambda^\delta{}_{\hat b}\,,
\end{equation}
where  $C^*_{\alpha\beta\gamma\delta}$ is the unique dual of the Weyl tensor given by
\begin{equation}\label{I7b} 
C^*_{\alpha\beta\gamma\delta}=\frac12 \eta^{\mu\nu}{}_{\alpha \beta}\,C_{\mu\nu\gamma\delta}\,,
\end{equation}
since the right and left duals of the Weyl  tensor coincide. Here, $\eta_{\mu\nu\rho\sigma}$ is the Levi-Civita tensor and in our convention, $\eta_{\hat 0 \hat 1 \hat 2 \hat 3}=1$, while $\eta_{\hat 0 \hat a \hat b \hat c}=\epsilon_{\hat a \hat b \hat c}$. Let us note that 
\begin{equation}\label{17c}
\mathcal{H}_{\hat a \hat b}= \frac12 \eta^{\mu\nu}{}_{\alpha \beta}\,C_{\mu\nu\gamma\delta}\,\lambda^\alpha{}_{\hat 0}\,\lambda^\beta{}_{\hat b}\, \lambda^\gamma{}_{\hat 0}\, \lambda^\delta{}_{\hat a}=\frac12 \eta^{\mu\nu}{}_{\hat 0 \hat b}\, C_{\mu\nu\hat 0 \hat a} = \frac12 C_{\hat 0 \hat a\hat c\hat d}\,\epsilon^{\hat c \hat d}{}_{\hat b}\,.
\end{equation}

Consider next a congruence of free test masses (``observers") following geodesics in a  gravitational field. We choose a reference observer in this
congruence and  set up a Fermi coordinate system along its world line. This 
amounts to constructing a local quasi-inertial system of coordinates
in the immediate neighborhood of the reference observer \cite{Syn}. 
Let $\lambda^\mu{}_{\hat \alpha}(\tau )$ be the natural
orthonormal tetrad frame that is parallel transported along the path of the reference observer as in display~\eqref{I3}. The quasi-inertial Fermi system with Fermi coordinates $X^{\hat \mu} =(\mathbb{T},{\bf X})$ is a natural geodesic
reference system along the world line of the observer and is based on the nonrotating orthonormal tetrad frame $\lambda^\mu{}_{\hat \alpha}$. Along the reference geodesic $\mathbb{T}=\tau$, ${\bf X}=0$ and 
$g_{\hat \mu \hat \nu}=\eta_{\hat \mu \hat \nu}$ by construction. The Fermi coordinates are
admissible within a cylindrical spacetime region around the world line of the reference observer of radius $\sim \mathcal{R}$, where $\mathcal{R}$ is the radius of curvature of spacetime~\cite{CM}.

The geodesic equation  in these Fermi coordinates is the equation of  motion of a free test particle in the congruence relative to the reference observer that is fixed at the spatial origin 
of Fermi coordinates.  It can be expressed as
\begin{equation}\label{I8} 
 \frac{d^2X^{\hat i}}{d\mathbb{T}^2}+R_{\hat 0\hat i\hat 0\hat j}X^{\hat j}+2\,R_{\hat i\hat k\hat j\hat 0}V^{\hat k}X^{\hat j}+\frac{2}{3}\,\left(3R_{\hat 0\hat k\hat j\hat 0}V^{\hat i}V^{\hat k}
+R_{\hat i\hat k\hat j\hat l}V^{\hat k}V^{\hat l}+ R_{\hat 0\hat k\hat j\hat l}V^{\hat i}V^{\hat k}V^{\hat l}\right) X^{\hat j}=0,
\end{equation}
which is valid to linear order in the separation ${\bf X}$. This geodesic 
deviation equation is a generalized Jacobi equation
\cite{CM} in which the rate of geodesic separation (i.e., the relative 
velocity of the test particle) ${\bf V}=d{\bf X}/d\mathbb{T}$ is
in general arbitrary ($|{\bf V}|<1$ at $\mathbf{X}=\mathbf{0}$).  It is clear from Eq.~\eqref{I8} that \emph{all} of the curvature components in Eq.~\eqref{I4} can be measured from a careful study of the motion of the test masses in the congruence relative to the fiducial observer. Neglecting terms  in relative velocity, 
 Eq.~\eqref{I8}  reduces to the Jacobi equation, namely, 
\begin{equation}\label{I9} 
\frac{d^2{X^{\hat i}}}{d\mathbb{T}^2} +\mathcal{K}^{\hat i}{}_{\hat j} X^{\hat j}=0\,,
\end{equation}
which is the relativistic analog of the Newtonian tidal equation given by Eq.~\eqref{I1}. In Eq.~\eqref{I9}, 
\begin{equation}\label{I10} 
\mathcal{K}_{\hat i \hat j}=R_{\hat 0 \hat i \hat 0 \hat j}\,,
\end{equation}
which is  an element of the  symmetric matrix $\mathcal{E}$ evaluated along the reference geodesic. This relativistic tidal matrix reduces in the nonrelativistic limit to the Newtonian tidal matrix~\eqref{I2}. 

The exterior gravitomagnetic field of the Earth has recently been measured directly via the Gravity Probe B (``GP-B") experiment~\cite{Ever}, which involved four superconducting gyroscopes and a telescope that were launched in 2004 into a polar orbit about the Earth aboard a drag-free satellite. The gravitomagnetic field of a rotating mass can also influence the relative (tidal) acceleration of nearby test particles via its contribution to the spacetime curvature. In 1980, Braginsky and Polnarev~\cite{BaPo} proposed an experiment to measure the relativistic rotation-dependent tidal acceleration of the Earth in a space platform in orbit around the Earth, since they claimed that such an approach could  
circumvent many of the difficulties associated with the GP-B experiment. However, Mashhoon and Theiss~\cite{MaTh} demonstrated that to measure the relativistic rotation-dependent tidal acceleration in a space platform, the local gyroscopes carried by the space platform must satisfy the same performance criteria as in the GP-B experiment. 

In future space experiments, it may be possible to combine the achievements of the GP-B with Paik's superconducting gravity gradiometer~\cite{Paik} in order to measure the tidal influence of the gravitomagnetic field using an orbiting platform~\cite{PMW, MPW}. The main purpose of this paper is to clarify the nature of the tidal matrix in such experiments.

\section{Gravity Gradiometry in Kerr Spacetime}

Let us first consider the exterior Kerr spacetime with the metric~\cite{Chandra}
\begin{equation}\label{K1}
-ds^2=-dt^2+\frac{\Sigma}{\Delta}dr^2+\Sigma\, d\theta^2 +(r^2+a^2)\sin^2\theta\, d\phi^2+\frac{2Mr}{\Sigma}(dt-a\sin^2\theta\, d\phi)^2\,,
\end{equation}
where $M$ is the mass of the gravitational source,  $a=J/M$ is the specific angular momentum of the source, $(t,r,\theta,\phi)$ are the standard Boyer--Lindquist coordinates and
\beq\label{K2}
\Sigma=r^2+a^2\cos^2\theta\,,\qquad \Delta=r^2-2Mr+a^2\,.
\eeq
In this paper, we consider a test mass $m$ and assume that its trajectory is a future directed timelike geodesic world line about a Kerr source.

The Kerr metric contains the gravitoelectric and gravitomagnetic potentials $\mathbb{U}$ and $\mathbb{V}$, which correspond to the mass and angular momentum of the source, respectively, and are  given by the dimensionless  quantities
\beq\label{K2a}
\mathbb{U}= \frac{GM}{c^2r}\,, \qquad \mathbb{V} = \frac{GJ}{c^3\,r^2}\,.
\eeq
For instance, in the case of the Earth with $r\approx R_{\oplus}=6.4\times 10^8$ cm, we have $\mathbb{U}_{\oplus} \approx 6\times 10^{-10}$ and 
 $\mathbb{V}_{\oplus} \approx 4\times 10^{-16}$. Therefore, for the exterior of the Earth
\beq\label{K2b}
\frac{a}{c\,r}= \frac{\mathbb{V}}{\mathbb{U}}\,
\eeq
is a quantity that is $<10^{-6}$. Furthermore, let us define  ratio $\rho(r)$  by
\begin{equation}\label{K2bA}
\rho(r) :=\frac{a}{c\,r\,\sqrt{\mathbb{U}}}\,,
\end{equation}
so that $\rho(r)$ is  \emph{independent of the speed of light $c$} as well as dimensionless. For the exterior of the Earth,
\begin{equation}\label{K2bB}
\rho(r) < \rho_{\oplus} \approx 3\times 10^{-2}\,.
\end{equation}
It turns out that $\rho(r)$ will play a significant role in the considerations of this paper. 

If we ignore, for the sake of simplicity, terms of order 
$(a/r)^2$ and higher in the treatment of the Kerr metric, Eq.~\eqref{K1} reduces to the Schwarzschild metric plus the Thirring--Lense term, namely, 
\begin{equation}\label{K2c}
-ds^2=-\left(1-\frac{2\,M}{r}\right)\,dt^2+\frac{dr^2}{1-\frac{2\,M}{r}}+r^2\, d\theta^2 +r^2\,\sin^2\theta\, d\phi^2- \frac{4\,a\,M}{r}\,\sin^2\theta\, dt\,d\phi\,.
\end{equation}
We recall that the exterior vacuum field of a spherically symmetric source in general relativity is static and  is uniquely given by the Schwarzschild metric.  Small deviations of the source from spherical symmetry can be treated via perturbations of the  Schwarzschild metric.  This general approach leads to the \emph{post-Schwarzschild approximation scheme.}  It will be employed later on in this paper  using the gravitational field given by metric~\eqref{K2c} for the treatment of the Mashhoon--Theiss effect~\cite{MaTh}, which appears in the gravitomagnetic part of the relativistic tidal matrix when neighboring test particles follow an inclined ``circular" orbit about a slowly rotating mass. It is important to emphasize that in Eq.~\eqref{K2c}, the mass of the gravitating source is taken into account to all orders, while the angular momentum of the source is taken into account only to linear order---see Section VII. The rest of the present section is devoted to a discussion of future directed timelike geodesic orbits and their parallel-propagated tetrad frames in the exterior Kerr spacetime.

\subsection{Circular Equatorial Geodesics}

Imagine a stable circular geodesic orbit of fixed radius $r_0$ in the equatorial $(x, y)$ plane with $\theta=\pi/2$. As is well known, such orbits exist from infinity all the way down to the last stable circular geodesic orbits $r_{\rm (LSO)\pm}$, which are  solutions of the equation 
\beq\label{K3}
1-\frac{6M}{r}\pm 8 a \sqrt{\frac{M}{r^3}}-3\frac{a^2}{r^2}=0\,.
\eeq
We use the convention that the upper (lower) sign refers to orbits where $m$ rotates in the same (opposite) sense as the source.
For $r<r_{\rm (LSO)\pm}$, there are unstable circular orbits that end at the null circular geodesic orbits given by
\beq\label{K4}
1-\frac{3M}{r}\pm 2 a \sqrt{\frac{M}{r^3}}=0\,.
\eeq

We define the Keplerian frequency $\omega_0$ for the circular orbits of radius $r_0$ under consideration here as
\beq\label{K5}
\omega_0^2 =\frac{M}{r_0^3}\,.
\eeq
The sign of $\omega_0$ would normally indicate the sense of the orbit; however, it is interesting to note that in Eqs.~\eqref{K3} and~\eqref{K4}, a prograde orbit becomes retrograde and vice versa when $a\to -a$. In the field of a central  rotating mass, an orbit can in general be either prograde or retrograde; therefore, it is natural to expect that certain orbital properties would depend upon $a\,\omega_0$, see below.  The reference world line is a geodesic; hence, 
\beq\label{K6}
t=\frac{1+a\,\omega_0}{N}\,\tau \,,\qquad \phi=\frac{\omega_0}{N}\,\tau+ \varphi_0\,,
\eeq
where $\tau$ is the proper time along the fiducial equatorial geodesic such that  $t=\tau=0$ at $\phi=\varphi_0$, $\varphi_0$ is a constant angle and
\beq\label{K7}
N=\sqrt{1-\frac{3M}{r_0}+2a\, \omega_0}\,.
\eeq
It is clear from a comparison of Eqs.~\eqref{K4} and~\eqref{K7} that $N=0$ in the limiting case of null circular orbits.  In connection with the timelike and azimuthal Killing vectors $\partial_t$ and $\partial_\phi$ of the background Kerr spacetime, the reference geodesic path has constants of motion, namely, specific energy $E$ and orbital angular momentum $\Phi$, respectively, given by
\beq\label{K8}
E=\frac{1}{N}\,\left( 1-\frac{2M}{r_0}+a\omega_0 \right)\,,\qquad
\Phi=\frac{r_0^2\,\omega_0}{N}\,\left(1-2a\,\omega_0 +\frac{a^2}{r_0^2}  \right)\,.
\eeq
Furthermore, Carter's constant $K$, associated with the Killing--Yano tensor of Kerr spacetime, is given for the circular geodesic orbit by~\cite{Marck}
\beq\label{K9}
K=(\Phi-a\,E)^2 = \frac{(r_0^2\,\omega_0-a)^2}{N^2}\,.
\eeq

\subsubsection{Tetrad Frame $\lambda^\mu{}_{\hat \alpha}$}

Next, we must determine $\lambda_{\hat{\alpha}}=\lambda^\mu{}_{\hat{\alpha}}\partial_\mu$, which is an orthonormal tetrad frame that undergoes parallel propagation along the reference geodesic world line with $\lambda^\mu{}_{\hat{0}}=dx^\mu/d\tau$. To this end,  let us first consider the natural tetrad frame field
$\Lambda_{\hat{\alpha}}=\Lambda^\mu{}_{\hat{\alpha}}\partial_\mu$ carried by the static observers in the exterior Kerr spacetime. In terms of the Boyer--Lindquist coordinates, the natural \emph{orthonormal} tetrad of the static observers \emph{in the equatorial plane}  along the  $(t,r,\theta,\phi)$ coordinate directions is given by
\begin{eqnarray}\label{K10}
\Lambda_{\hat{0}}&=&\frac{1}{A}\partial_t \,,\qquad \Lambda_{\hat{1}}=\frac{\sqrt{\Delta}}{r}\partial_r\nonumber\\
\Lambda_{\hat{2}}&=&\frac{1}{r}\partial_\theta \,,\qquad \Lambda_{\hat{3}}=-\frac{2Ma}{r\,A\sqrt{\Delta}}\partial_t +\frac{A}{\sqrt{\Delta}}\partial_\phi\,,
\end{eqnarray}
where 
\beq\label{K11}
A=\sqrt{1-\frac{2M}{r}}\,.
\eeq
Let us now subject this tetrad frame field, \emph{restricted to be along the fiducial orbit at $r=r_0$}, to a Lorentz boost with speed $\tilde \beta$, $\Lambda^\mu{}_{\hat{\alpha}}\mapsto \tilde \lambda^\mu{}_{\hat{\alpha}}$, such that 
$\tilde \lambda_{\hat{0}}=\lambda_{\hat{0}}$ is the unit vector tangent to the reference world line. In this way, we get an orthonormal tetrad frame along the fiducial orbit given by
\begin{equation}\label{K12}
\tilde \lambda_{\hat{0}}=\tilde \gamma\, [\Lambda_{\hat{0}}+\tilde \beta\, \Lambda_{\hat{3}}]\,,\qquad  \tilde \lambda_{\hat{1}}=\Lambda_{\hat{1}}\,,
\end{equation}
\begin{equation}\label{K13}
\tilde \lambda_{\hat{2}}=\Lambda_{\hat{2}}\,,\qquad \tilde \lambda_{\hat{3}}=\tilde \gamma\, [\Lambda_{\hat{3}}+\tilde \beta\, \Lambda_{\hat{0}}]\,,
\end{equation}
where $(\tilde \beta, \tilde \gamma)$ is a Lorentz pair. That is, $\tilde \gamma$ is the Lorentz factor corresponding to speed $\tilde \beta$. This Lorentz pair can be determined from $\tilde \lambda_{\hat{0}}=\lambda_{\hat{0}}$; hence, we find 
\beq\label{K14}
\tilde \beta=\frac{\sqrt{\Delta_0} \,\omega_0}{EN}\,,\qquad \tilde \gamma =\frac{E}{A_0}\,,
\eeq
where $\Delta_0 := \Delta(r=r_0)$ and, similarly, $A_0:=A(r=r_0)$, namely,
\begin{equation}\label{K15}
\Delta_0= r_0^2\left(1-\frac{2M}{r_0}+\frac{a^2}{r_0^2}\right)\,, \qquad  A_0=\sqrt{1-\frac{2M}{r_0}}\,.
\end{equation}
 It follows from display~\eqref{K14} that, for $a>0$ as in Figure 1,  $\tilde \beta$ is positive (negative) for prograde (retrograde) orbits; moreover, $\tilde \gamma$ diverges at the null orbits $(N=0)$.
 

\begin{figure}\label{FIG. 1}
\includegraphics[scale=0.5]{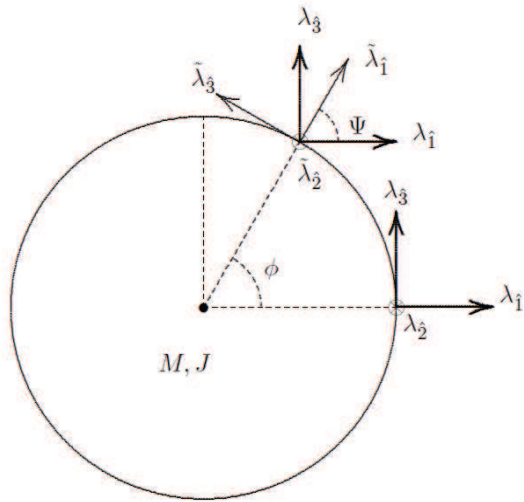}
\caption{This schematic plot depicts the construction of a natural spatial triad that is parallel propagated along a stable circular geodesic orbit in the equatorial $(x, y)$ plane of the exterior Kerr spacetime. The result coincides in this case with the general frame constructed by Marck for an arbitrary geodesic world line in the exterior Kerr spacetime. That is, Marck's construction can be viewed as a simple generalization of the method illustrated in this figure.}
\end{figure}


It is clear from our construction of the boosted tetrad that the spatial triad $\tilde \lambda_{\hat{a}}$, $a=1,2,3$, points primarily along the spherical polar coordinate directions, namely,  the radial, normal and tangential directions with respect to the circular orbit, see Figure 1. Intuitively, to have a frame that is parallel propagated, we need to rotate the boosted frame back, as illustrated in Figure 1. That is, we must solve the parallel transport equations for the angle $\Psi$ such that the resulting tetrad $\lambda_{\hat{\alpha}}$ would be parallel propagated along the orbit. Thus we have
\begin{eqnarray}\label{K16}
\lambda_{\hat{0}}&=&\tilde \lambda_{\hat{0}}\,,\qquad \lambda_{\hat{1}}=\tilde \lambda_{\hat{1}} \cos \Psi - \tilde \lambda_{\hat{3}} \sin\Psi\,,\nonumber\\
\lambda_{\hat{2}}&=&\tilde \lambda_{\hat{2}}\,,\qquad \lambda_{\hat{3}}=\tilde \lambda_{\hat{1}} \sin \Psi + \tilde \lambda_{\hat{3}} \cos\Psi\,.
\end{eqnarray}
The covariant derivative of $\lambda_{\hat{a}}$ vanishes along the orbit; hence,  we find 
\beq\label{K17}
\Psi =\omega_0 \tau\,,
\eeq
where we have assumed that $\Psi=0$ at $\tau=0$. Any other parallel-transported spatial frame along the orbit can be determined from $\lambda_{\hat{a}}$ by a \emph{constant} rotation involving, for instance, three constant Euler angles. 

The difference between the angles $\phi$ and $\Psi$ in Figure 1 is due to the precession of an ideal gyro in the field of a rotating mass. Indeed, 
\beq\label{K18}
\phi -\Psi=\left(\frac{1}{N}-1  \right)\Psi\,,
\eeq
which, to first order in $M/r_0\ll 1$ and $a/r_0\ll 1$ can be written as
\beq\label{K19}
\phi-\Psi  \approx \left(\frac32 \frac{M}{r_0}\,\omega_0 -\frac{J}{r_0^3}  \right)\,\tau\,.
\eeq
In the equatorial plane, the difference between these angles is due to a combination of prograde geodetic and retrograde gravitomagnetic precessions. 

The end result of our approach to the construction of  the frame $\lambda_{\hat{\alpha}}$ along the circular geodesic orbit of test mass $m$ in the equatorial plane of exterior Kerr geometry can be given in $(t,r,\theta, \phi)$ coordinates as
\begin{eqnarray}\label{K20}
\lambda_{\hat{0}}&=&\frac{1}{N}\left[(1+a\,\omega_0)\,\partial_t +\omega_0\, \partial_\phi\right]\,,\nonumber\\
{}\lambda_{\hat{1}}&=&\frac{1}{\sqrt{\Delta_0}}\left[
-\Phi\,\sin (\omega_0 \tau)\, \partial_t +\frac{\Delta_0}{r_0}\cos (\omega_0 \tau)\, \partial_r - E\sin (\omega_0 \tau)\,  \partial_\phi
\right]\,,\nonumber\\
{}\lambda_{\hat{2}}&=&\frac{1}{r_0}\,\partial_\theta\,,\nonumber\\
{}\lambda_{\hat{3}}&=&\frac{1}{\sqrt{\Delta_0}}\left[\Phi \cos (\omega_0 \tau)\, \partial_t +\frac{\Delta_0}{r_0}\sin (\omega_0 \tau)\, \partial_r + E\cos (\omega_0 \tau)\,  \partial_\phi\right]\,.
\end{eqnarray}
It is important to recognize that the parallel-propagated spatial frame is unique up to a \emph{constant} spatial rotation. 

\subsubsection{Measured Components of Curvature}

The projection of the Weyl tensor on  the frame $\lambda_{\hat{\alpha}}$ along the circular  orbit in the equatorial plane of the source-free  exterior region of spacetime under consideration may be expressed in the standard manner as 
\beq\label{K21}
\mathcal {E}= \omega_0^2\, \left[
\begin{array}{ccc}
k_1 & 0 & k'\cr
0 & k_2 & 0\cr
k' & 0 & k_3\cr
\end{array}
\right]\,,\qquad
\mathcal {H}= \omega_0^2\, \left[
\begin{array}{ccc}
0 & h & 0\cr
h & 0 & h'\cr
0 & h' & 0\cr
\end{array}
\right]\,,
\eeq
where ${\mathcal E}$ and ${\mathcal H}$ are $3\times 3$ symmetric and traceless matrices containing respectively the gravitoelectric and gravitomagnetic components of the Weyl curvature tensor. Here, $k_2=-(k_1+k_3)$ is constant and is given by $k_2=3\gamma^2-2$, while
\begin{eqnarray}\label{K22}
k_1&=&1-3\gamma^2 \cos^2(\omega_0 \tau)\,,\qquad k_3=1-3\gamma^2 \sin^2(\omega_0 \tau)\,,\qquad k'=-3\gamma^2 \sin (\omega_0 \tau)\, \cos (\omega_0 \tau)\,,\nonumber\\
h&=& -3\gamma^2\beta \cos (\omega_0 \tau)\,,\qquad h'=-3\gamma^2 \beta \sin (\omega_0 \tau)\,.
\end{eqnarray}
In these equations, $(\beta, \gamma)$ is a new Lorentz pair given by
\beq\label{K23}
\beta=\frac{r_0^2\,\omega_0-a}{\sqrt{\Delta_0}}\,,\qquad \gamma =\frac{\sqrt{\Delta_0}}{r_0\,N}\,.
\eeq
We note that for the null orbits $(N=0)$, $\gamma$ diverges and as $a \to 0$,  $(\beta, \gamma) \to (\tilde \beta, \tilde \gamma)$ in the Schwarzschild limit. To linear order in $a/r_0 \ll 1$, we can write
\beq\label{K23A}
\beta=\sqrt{\frac{u}{1-2u}}\,\left(1-\frac{a\,\omega_0}{u}\right)+ \mathcal{O}\left(\frac{a^2}{r_0^2}\right)\,
\eeq
and
\beq\label{K23B}
\gamma=\sqrt{\frac{1-2u}{1-3u}}\,\left(1-\frac{a\,\omega_0}{1-3u}\right)+ \mathcal{O}\left(\frac{a^2}{r_0^2}\right)\,,
\eeq
where $u$ is the dimensionless quantity
\beq\label{K23C}
u:=\frac{M}{r_0}\,,
\eeq
which is much less than unity for the practical considerations that have motivated this work. In particular, for orbits around the Earth with $\omega_0 >0$, $u < \mathbb{U}_{\oplus}$ and $\beta > 0$, since $r_0^2\,\omega_0>a$. This relation follows from  $\rho(r_0)=a\,\omega_0/(c^2\,u) < 3 \times 10^{-2}$ by Eq.~\eqref{K2bB}. 

The measured components of the curvature are \emph{periodic} in this case; in fact, the gravitoelectric part (i.e., the relativistic tidal matrix) consists of constant terms plus Fourier terms that vary with frequency $2\,\omega_0$ with respect to proper time $\tau$, while the gravitomagnetic terms are all off-diagonal and vary with frequency $\omega_0$ with respect to $\tau$. It is intuitively  clear that the \emph{periodic} nature of the measured curvature components would be preserved under a constant rotation of the spatial frame. 

It is interesting to observe that when $0\le a\le M$, $\beta \in [-\frac12, \frac12]$ for the circular orbits under consideration; in fact, $\beta$ is positive (negative) for a prograde (retrograde) orbit and $|\beta|\sim \sqrt{u}$ far away from the source $(r_0\gg 2M)$. Therefore, $\beta \to 0$ as $r_0 \to \infty$; hence $|\beta|$ monotonically decreases from $1/2$ to zero as $r_0$ increases from $r_{\rm (LSO)\pm}$ to infinity.  On the other hand, when $a>M$, $\beta$ is always negative for retrograde orbits, but is not always positive for prograde orbits; indeed, $\beta$ vanishes for a prograde orbit of radius $r_0=a^2/M$.

Our results for the relativistic tidal matrix $\mathcal{E}$ are in agreement with the work of Marck~\cite{Marck}.

\subsubsection{Components of Curvature Projected on $\tilde \lambda_{\hat \alpha}$}

To gain further insight into the nature of the measured components of curvature, it is interesting to study the projection of the Weyl tensor on the frame $\tilde \lambda_{\hat \alpha}$ that is rotated by angle $\Psi$ relative to frame $\lambda_{\hat \alpha}$ as in Eq.~\eqref{K16}. In this case, the measured components of the curvature tensor are
\beq\label{K23a}
\left[
\begin{array}{cc}
\tilde{\mathcal{E}} & \tilde{\mathcal{H}}\cr
\tilde{\mathcal{H}} & -\tilde{\mathcal{E}}\cr 
\end{array}
\right]
\eeq
where $(\tilde{\mathcal{E}}, \tilde{\mathcal{H}})$ are related to  $(\mathcal{E}, \mathcal{H})$  via rotation~\eqref{K16}. Let us denote the transformation of the triad in Eq.~\eqref{K16} by the rotation matrix $S$, $S^{\dagger}=S^{-1}$, where
\beq\label{K23b}
S= \left[
\begin{array}{ccc}
\cos \Psi & 0 & -\sin \Psi\cr
0 & 1 & 0\cr
\sin \Psi & 0 & \cos \Psi\cr
\end{array}
\right]\,.
\eeq
Under such a rotation, it is straightforward to show that the gravitoelectric part (i.e., the relativistic  tidal matrix) and the gravitomagnetic part of the Weyl tensor undergo a similarity transformation, namely,
\begin{equation}\label{K23c}
\mathcal{E}=S\,\tilde{\mathcal{E}}\,S^{-1}\,, \qquad \mathcal{H}=S\,\tilde{\mathcal{H}}\,S^{-1}\,,
\end{equation}
which can be expressed in components as
\begin{eqnarray}\label{K23d}
\frac{1}{2}\,(\mathcal{E}_{11}-\mathcal{E}_{33})&=&\frac{1}{2}\,(\tilde{\mathcal{E}}_{11}-\tilde{\mathcal{E}}_{33})\,\cos 2\Psi - \tilde{\mathcal{E}}_{13}\,\sin 2\Psi\,,\nonumber\\
\mathcal{E}_{13}&=&\tilde{\mathcal{E}}_{13}\,\cos 2\Psi+\frac{1}{2}\,(\tilde{\mathcal{E}}_{11}-\tilde{\mathcal{E}}_{33})\,\sin 2\Psi\,, \nonumber\\
\mathcal{E}_{22}&=&\tilde{\mathcal{E}}_{22}\,,\nonumber\\
\mathcal{E}_{12}&=&\tilde{\mathcal{E}}_{12} \cos \Psi -\tilde{\mathcal{E}}_{23} \sin \Psi\,, \nonumber\\
\mathcal{E}_{23}&=&\tilde{\mathcal{E}}_{23} \cos \Psi +\tilde{\mathcal{E}}_{12} \sin \Psi\,,
\end{eqnarray}
 while for $\mathcal{H}$ we have 
\begin{eqnarray}\label{K23e}
\frac{1}{2}\,(\mathcal{H}_{11}-\mathcal{H}_{33})&=&\frac{1}{2}\,(\tilde{\mathcal{H}}_{11}-\tilde{\mathcal{H}}_{33})\,\cos 2\Psi - \tilde{\mathcal{H}}_{13}\,\sin 2\Psi\,,\nonumber\\
\mathcal{H}_{13}&=&\tilde{\mathcal{H}}_{13}\,\cos 2\Psi+\frac{1}{2}\,(\tilde{\mathcal{H}}_{11}-\tilde{\mathcal{H}}_{33})\,\sin 2\Psi\,, \nonumber\\
\mathcal{H}_{22}&=&\tilde{\mathcal{H}}_{22}\,,\nonumber\\
\mathcal{H}_{12}&=&\tilde{\mathcal{H}}_{12} \cos \Psi -\tilde{\mathcal{H}}_{23} \sin \Psi\,, \nonumber\\
\mathcal{H}_{23}&=&\tilde{\mathcal{H}}_{23} \cos \Psi +\tilde{\mathcal{H}}_{12} \sin \Psi\,.
\end{eqnarray}
It is clear that transformations~\eqref{K23d} and~\eqref{K23e} can be simply reversed if $\Psi$ is replaced by $-\Psi$. 

For the circular equatorial geodesic under consideration, we find from the inverse of the transformations~\eqref{K23d}--\eqref{K23e} with $\Psi=\omega_0\, \tau$ that 
\beq\label{K23f}
\tilde{\mathcal{E}}= \omega_0^2\, \left[
\begin{array}{ccc}
1-3\gamma^2 & 0 & 0\cr
0 & -2+3\gamma^2 & 0\cr
0 & 0 & 1\cr
\end{array}
\right]\,,\qquad
\tilde{\mathcal{H}}= \omega_0^2\, \left[
\begin{array}{ccc}
0 & -3\gamma^2\,\beta & 0\cr
-3\gamma^2\,\beta & 0 & 0\cr
0 & 0 & 0\cr
\end{array}
\right]\,.
\eeq
We recover the expected \emph{diagonal} Newtonian tides in the gravitoelectric components of the curvature tensor as $(\beta, \gamma) \to (0, 1)$ for $r_0 \to \infty$. 
These results are clearly consistent with the nature of the frame under consideration, which corresponds to the radial, normal and tangential directions along the circular geodesic orbit. In this natural frame, we note  the presence of only \emph{off-diagonal} gravitomagnetic components of the curvature tensor.

\subsection{Marck's Frame for an Arbitrary Geodesic Orbit}

We now turn our attention to an arbitrary geodesic world line in the exterior Kerr spacetime. As is well known, the geodesic equation can be integrated in this case and the first integrals of the motion are given by
\begin{equation}\label{K24}
\Delta\,\Sigma\,\frac{dt}{d\tau}=[(r^2+a^2)\Sigma + 2 M r a^2 \sin^2\theta]E-2Mra\Phi\,,
\end{equation}
\begin{equation}\label{K25}
\Sigma^2\,\Big(\frac{dr}{d\tau}\Big)^2=[(r^2+a^2)E-a\Phi]^2-\Delta(r^2+K)\,, 
\end{equation}
\begin{equation}\label{K26}
\Sigma^2\,\Big(\frac{d\theta}{d\tau}\Big)^2=K-a^2\cos^2\theta-\Big(aE\sin \theta-\frac{\Phi}{\sin \theta}\Big)^2\,,
\end{equation}
\begin{equation}\label{K27}
\Delta\,\Sigma\,\frac{d\phi}{d\tau}=2MraE+(\Sigma-2Mr)\,\frac{\Phi}{\sin^2 \theta}\,.
\end{equation}

\subsubsection{Marck's Tetrad Frame $\lambda^\mu{}_{\hat \alpha}$}

Marck has shown how the procedure we followed for a circular orbit can be generalized to an arbitrary geodesic in the exterior Kerr spacetime~\cite{Marck}. Using Kerr's Killing--Yano tensor $f_{\mu\nu}=-f_{\nu \mu}$, whose nonvanishing components are given by
\beq
\label{KY}
f_{tr}=-a\cos\theta\,,\qquad f_{t\theta}=ar \sin \theta \,,\qquad f_{r\phi}=-a^2\cos \theta \sin^2 \theta \,,\qquad f_{\theta\phi}=(r^2+a^2)r\sin \theta \,,
\eeq
and which satisfies
\begin{equation}\label{K28}
\nabla_\rho f_{\mu \nu} + \nabla_\nu f_{\mu \rho}=0\,,
\end{equation}
together with $\lambda^\mu {}_{\hat 0}=dx^\mu/d\tau$, the 4-velocity of an arbitrary geodesic, one can construct $L_\mu=f_{\mu\nu}\lambda^\nu{}_{\hat 0}$. This vector is then orthogonal to $\lambda_{\hat 0}$  and is parallel propagated along the geodesic orbit.  Carter's constant is defined by $K:=L_\mu\,L^\mu$, so that $K$ is positive by construction, since $L_\mu$ is a spacelike vector. Moreover, $K$ is  constant along the orbit. In this way, Marck  obtained
\begin{equation}\label{K29}
\lambda^\mu{}_{\hat 2}=\frac{1}{\sqrt{K}}f^\mu{}_\nu \lambda^\nu{}_{\hat 0}\,.
\end{equation}
Next, Marck found by inspection $\tilde \lambda^\mu{}_{\hat 1}$ and $\tilde \lambda^\mu{}_{\hat 3}$, which together with $\tilde \lambda^\mu{}_{\hat 2}=\lambda^\mu{}_{\hat 2}$ and $\tilde \lambda^\mu{}_{\hat 0}=\lambda^\mu{}_{\hat 0}$ form an orthonormal tetrad frame. However, to get a frame that is parallel propagated, one must find $\Psi$ such that
\beq\label{K30}
\lambda_{\hat 1}= \tilde \lambda_{\hat 1} \cos \Psi - \tilde \lambda_{\hat 3} \sin \Psi\,,\qquad
\lambda_{\hat 3}= \tilde \lambda_{\hat 1} \sin \Psi + \tilde \lambda_{\hat 3} \cos \Psi\,.
\eeq
The result is~\cite{Marck}
\beq\label{K31}
\frac{d\Psi}{d\tau}=\frac{\sqrt{K}}{\Sigma}\left[ \frac{(r^2+a^2)E-a\Phi}{r^2+K}+a\, \frac{\Phi-aE\sin^2\theta}{K-a^2\cos^2\theta} \right]\,,
\eeq
which must be integrated along the orbit to determine $\Psi$. Explicitly, for a general orbit given by Eqs.~\eqref{K24}--\eqref{K27}, Marck's tetrad coframe  can be obtained from
\beq\label{K32}
\lambda_{\hat 0}= \lambda_{\mu \hat 0}\,dx^\mu =-E\, dt +\frac{\Sigma}{\Delta}\dot r\, dr +\Sigma \dot \theta \, d\theta +\Phi \,d\phi\,,
\eeq
where an overdot denotes differentiation with respect to proper time $\tau$, and
\beq\label{K33}
\tilde \lambda_{\hat 1}=\frac{1}{\sqrt{K}}\,\mathcal {U}_\alpha\, dx^\alpha\,,\qquad
\tilde \lambda_{\hat 2}=\lambda_{\hat 2}=\frac{1}{\sqrt{K}}\,\mathcal{V}_\alpha \,dx^\alpha\,,\qquad
\tilde \lambda_{\hat 3}=\frac{1}{\Sigma}\mathcal{W}_\alpha\, dx^\alpha\,.
\eeq
To avoid confusion here, we mention that we denote frame vectors and coframe 1-forms using the same symbol, namely, $\tilde \lambda_{\hat \alpha}$. We find, using the results given by Marck~\cite{Marck}, that
\begin{eqnarray}\label{K34}
\nonumber \mathcal {U}_0 &=&-\hat \alpha\, r \,\dot r -\hat \beta\, a^2 \sin \theta \cos \theta\, \dot \theta\,, \\
\nonumber \mathcal {U}_1 &=& \hat \alpha\, \frac{r}{\Delta}[(r^2+a^2)E-a\Phi]\,, \\
\nonumber \mathcal {U}_2 &=& \hat \beta\, a \cos\theta \left(aE\sin \theta-\frac{\Phi}{\sin \theta}  \right)\,, \\
\mathcal {U}_3 &=& \hat \alpha\, a\sin^2\theta \,r \,\dot r +\hat \beta\, a(r^2+a^2) \sin \theta \cos\theta \,\dot \theta\,,
\end{eqnarray}
\begin{eqnarray}\label{K35}
\nonumber \mathcal{V}_0 &=& -a \cos \theta \,\dot r + a r \sin \theta \,\dot \theta \,, \\
\nonumber \mathcal{V}_1 &=& \frac{a\cos\theta}{\Delta}[(r^2+a^2)E-a\Phi]\,, \\
\nonumber \mathcal{V}_2 &=& -r \left(aE\sin \theta -\frac{\Phi}{\sin\theta}  \right)\,, \\
\mathcal{V}_3 &=& a^2 \cos \theta \sin^2\theta\, \dot r -r(r^2+a^2) \sin\theta\, \dot \theta\,
\end{eqnarray}
and
\begin{eqnarray}\label{K36}
\nonumber \mathcal{W}_0 &=& -\hat \alpha\, [(r^2+a^2)E-a\Phi]+\hat \beta\, a (aE\sin^2\theta -\Phi)\,, \\
\nonumber \mathcal{W}_1 &=& \hat \alpha\, \frac{\Sigma^2}{\Delta}\,\dot r\,, \\
\nonumber  \mathcal{W}_2 &=& \hat \beta \,\Sigma^2 \,\dot \theta\,,  \\
\mathcal{W}_3 &=& \hat \alpha\, a \sin^2\theta\, [(r^2+a^2)E-a\Phi]-\hat \beta\, (r^2+a^2)(aE\sin^2\theta-\Phi)\,,
\end{eqnarray}
where
\beq \label{K37}
\hat \alpha =\sqrt{\frac{K-a^2\cos^2\theta}{K+r^2}}\,,\qquad \hat \beta =\frac{1}{\hat \alpha}\,.
\eeq
For a recent discussion of Marck's frame, see Ref.~\cite{Bini:2016iym}.

\section{Tilted Spherical Orbit About a Slowly Rotating Mass}

We now wish to work out Marck's tetrad system for the  tilted ``circular" Keplerian orbit of a test mass $m$ about a slowly rotating astronomical mass $M$.  Henceforward, the orbit will be assumed to have a positive Keplerian frequency $\omega_0>0$. We will do this calculation in several steps, starting with a circular orbit in the exterior Schwarzschild spacetime.

\subsection{Equatorial Circular Orbit with $a=0$}

Let us return to the stable circular orbit in the equatorial $(x, y)$ plane of Kerr spacetime and turn off the rotation of the source, i.e., we set $J=0$. Then, the Schwarzschild circular orbit with $\omega_0>0$ is given by
\begin{equation}\label{S1}
t=\frac{\omega}{\omega_0}\,\tau\,, \qquad r=r_0\,, \qquad \theta=\frac{\pi}{2}\,, \qquad 
\phi=\omega\, \tau + \varphi_0\,.
\end{equation}
Here, we have introduced
\begin{equation}\label{S2}
\omega:=\frac{\omega_0}{N_0}\,, \qquad N_0:=\sqrt{1-3\frac{M}{r_0}}\,.
\end{equation}
Moreover, for the orbit under consideration here, the specific energy $E_0$, specific orbital angular momentum $\Phi_0$ and Carter's constant $K_0$ are now given by
\begin{equation}\label{S3}
E_0=\frac{A_0^2}{N_0}\,, \qquad \Phi_0=r_0^2\,\omega\,, \qquad K_0=\Phi_0^2\,.
\end{equation}

\subsection{Tilted Circular Orbit with $a=0$}

Because of spherical symmetry we can have an arbitrary stable circular Keplerian orbit around a Schwarzschild source. We first need to choose such an orbit and later perturb it to linear order in $a=J/M$. 
To this end, let us consider the transformation from the background space to the tilted orbital plane. We first consider a rotation about the $z$ axis by an angle $\psi$ 
\begin{eqnarray}\label{S4}
\nonumber x&=& x' \cos\psi-y' \sin \psi\,, \\
\nonumber y&=& x' \sin\psi+y' \cos \psi\,,  \\
z&=& z'\,. 
\end{eqnarray}
Next, we rotate about $x'$ axis by the tilt angle $\alpha$
\begin{eqnarray}\label{S5}
\nonumber x'&=& x'' \,, \\
\nonumber y'&=& y'' \cos\alpha-z'' \sin \alpha\,, \\
z'&=& y'' \sin\alpha +z'' \cos \alpha\,.
\end{eqnarray}
A general rotation of spatial axes would involve three rotation angles.  For an arbitrary orbit, we 
therefore need another rotation about $z''$ axis; however, such a rotation is redundant as the orbit is circular in the $(x'', y'')$ plane.
Therefore, we write the parametric equations of the orbit as
\beq\label{S6}
x''=r_0 \cos (\omega\tau +\eta_0)\,,\quad y''=r_0 \sin (\omega\tau +\eta_0)\,,\quad z''=0\,,
\eeq 
where $r_0$ is the radius of the circular orbit and $\eta_0$ is a constant angle. 

It proves useful to define $\eta$,
\begin{equation}\label{S7}
\eta:= \omega\tau +\eta_0\,.
\end{equation}
Transforming back to the coordinates $(x',y',z')$, we have for the parametric equations of the orbit
\beq\label{S8}
x'=r_0 \cos \eta\,,\qquad y'=r_0 \cos\alpha\, \sin \eta\,,\qquad z'=r_0 \sin \alpha \,\sin \eta\,.
\eeq
Finally, in terms of $(x,y,z)$ coordinates we find 
\begin{eqnarray}\label{S9}
x&=& r_0 (\cos\psi \cos \eta-\cos\alpha \sin\psi \sin \eta)\,,\nonumber\\
y&=& r_0 (\sin\psi \cos \eta+\cos\alpha \cos\psi \sin \eta)\,,\nonumber\\
z&=& r_0\sin \alpha \sin \eta\,.
\end{eqnarray}
In polar coordinates $x=r_0 \sin \theta\,\cos\phi$, $y=r_0 \sin\theta \,\sin\phi$ and $z=r_0\cos\theta$, the parametric equations of the orbit can be  summarized as
\beq\label{S10}
\cos\theta = \sin \alpha\, \sin \eta\,,\qquad 
\tan\phi = \frac{\sin\psi\,\cos \eta+\cos\alpha\, \cos\psi \,\sin \eta}{\cos\psi\,\cos \eta-\cos\alpha\, \sin\psi\, \sin \eta}\,.
\eeq
In order to recover Eq.~\eqref{S1} for $\alpha=0$, we set
\begin{equation}\label{S11}
\psi=\varphi_0-\eta_0\,.
\end{equation}
Equations~\eqref{S9} simplify in the case of small inclination; that is, for $\alpha \ll 1$,
\begin{equation}\label{S12}
x= r_0  \cos(\omega\tau +\varphi_0)\,,\qquad
y= r_0  \sin(\omega\tau +\varphi_0)\,,\qquad
z= r_0\,  \alpha\, \sin \eta\,,
\end{equation}
so that in polar coordinates we have to linear order in $\alpha$
\beq\label{S13}
r=r_0\,,\quad \theta=\frac{\pi}{2}-\alpha\, \sin \eta\,,\quad \phi=\omega \tau+\varphi_0 \,.
\eeq

Let us now return to the general case and note that in display~\eqref{S10}, we can write
\beq\label{S14}
 \tan\phi = \frac{\tan\psi +\cos\alpha\, \tan\eta}{1-\cos\alpha\, \tan\eta\, \tan \psi}\,.
\eeq
It is useful to define $\chi$ such that 
\beq\label{S15}
 \tan\chi = \cos\alpha\, \tan\eta\,.
\eeq
Then, it follows from Eq.~\eqref{S14} that $\phi=\psi + \chi$. On the other hand, we know from Eq.~\eqref{S11} that $\psi=\varphi_0-\eta_0$. Putting all these results together, we conclude that the \emph{general tilted circular geodesic orbit in Schwarzschild spacetime} is given by
\beq\label{S16}
t=\frac{\omega}{\omega_0}\,\tau\,,\qquad r=r_0\,,\qquad \theta =\arccos(\sin \alpha \,\sin \eta)\,,\qquad \phi=\arctan(\cos\alpha\, \tan \eta )+\varphi_0 -\eta_0\,,
\eeq
where $\eta=\omega\,\tau +\eta_0$. Next, substituting Eq.~\eqref{S16} in the geodesic Eqs.~\eqref{K24}--\eqref{K27}, we find the generalization of Eq.~\eqref{S3} for the general tilted circular orbit, namely,
\begin{equation}\label{S17}
E_0=\frac{A_0^2}{N_0}\,, \qquad \Phi_0=r_0^2\,\omega\, \cos\alpha\,, \qquad K_0=r_0^4\,\omega^2\,.
\end{equation}
This orbit is depicted in Figure 2. Before we leave Schwarzschild spacetime, it is useful to collect here some formulas for future reference; that is, $u=M/r_0$,
\begin{equation}\label{S18}
A_0=\sqrt{1-2u}\,, \qquad N_0=\sqrt{1-3u}\,, \qquad r_0\, \omega_0 =\sqrt{u}\,. 
\end{equation}


\begin{figure}\label{FIG. 2}
\includegraphics[scale=0.5]{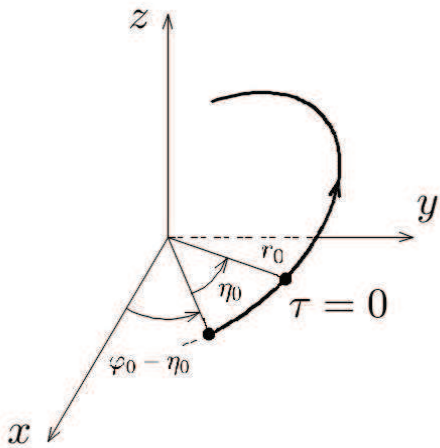}
\caption{Schematic  depiction of  the tilted circular orbit  followed by a test mass $m$  around a Schwarzschild source of mass $M\gg m$. The inclination of the orbit is given by the angle $\alpha$. Moreover, $\eta$ is the angular position of $m$ in the orbital plane  measured from the line of the ascending node. At $\tau=0$, $m$ has spherical  polar coordinates $(r_0, \theta_0, \phi_0)$, where $\cos \theta_0 = \sin \alpha \,\sin \eta_0$ and $\phi_0= \arctan(\cos \alpha \, \tan \eta_0) +\varphi_0-\eta_0$.}
\end{figure}


\subsection{Tilted ``Circular" Orbit to Linear Order in $a=J/M$}

Let us now turn on rotation, but only to first order in the small dimensionless parameter $a/r_0$. It turns out that at this order of approximation the geodesic equation allows $r=r_0$ to remain constant, but the orbit is no longer planar. Thus we have a tilted spherical orbit in a slowly rotating Kerr spacetime. Indeed, regular spherical orbits in Kerr spacetime have been studied in detail~\cite{Wilk}; however, the orbits under consideration here are \emph{tilted spherical orbits}. In practice, the tilted spherical orbit is in effect ``circular", as it tends to stay rather close to a circular orbit. 

To simplify the analysis, let us define functions $\bar\theta(\tau)$ and $\bar \phi (\tau)$ as follows:
\beq\label{J1}
\bar\theta(\tau) =\arccos(\sin \alpha \,\sin \eta)\,, \qquad \bar \phi(\tau)=\arctan(\cos\alpha\, \tan \eta )+\varphi_0 -\eta_0\,,
\eeq
where, as before, $\eta=\omega\,\tau +\eta_0$. We note that $\cos \bar \theta =\sin \alpha \sin \eta$ and since $\bar \theta:0\to \pi$, we have $\sin\bar\theta \ge 0$ and 
\beq\label{J2}
\sin \bar \theta =\sqrt{1-\sin^2\alpha \sin^2\eta}=\sqrt{\cos^2\eta +\cos^2\alpha\,\sin^2\eta}\,.
\eeq
Moreover, 
\beq\label{J3}
 \frac{d\bar \theta}{d\tau}=-\frac{\omega\,\sin\alpha\, \cos \eta}{\sin \bar \theta}\,,\qquad \frac{d\bar \phi}{d\tau}=\frac{\omega\, \cos \alpha}{\sin^2\bar \theta}\,.
\eeq
Then, the perturbed ``circular" orbit to linear order in $a$ is assumed to be of the form
\beq\label{J4}
t=\frac{\omega}{\omega_0}\,\tau+ a\,T(\tau)\,,\qquad r=r_0\,,\qquad \theta =\bar\theta(\tau)+a\, \Theta(\tau)\,,\qquad \phi=\bar \phi(\tau)+a\, F(\tau)\,,
\eeq
where $T$, $\Theta$ and $F$ are to be determined from the solution of the geodesic equation in Kerr spacetime to linear order in $a/r_0$ such that at $\tau=0$,  $T(0)=\Theta(0)=F(0)=0$. It follows from a detailed examination of the geodesic equation that 
\beq\label{J5}
T =-3 \frac{u\,\cos\alpha}{N_0^2}\,\omega\,\tau\,,
\eeq
\beq\label{J6}
\Theta =3\,\Big(\frac{A_0}{N_0}\Big)^2\,\omega_0\,\sin \alpha\, \cos \alpha\,\frac{\omega\,\tau\, \cos \eta}{\sin \bar \theta}\,,
\eeq
\beq\label{J7}
F = 2\, \omega_0 \,\omega\,\tau - 3\,\Big(\frac{A_0}{N_0}\Big)^2\,\omega_0\, \cos^2 \alpha\,\frac{\omega\,\tau}{\sin^2 \bar \theta}\,.
\eeq
It is useful to write the orbital equations as 
\begin{eqnarray}\label{J8}
t&=& \frac{\omega}{\omega_0}\tau -a (3\omega^3 r_0^2\cos\alpha)\tau\,,  \nonumber \\
r &=&  r_0\,, \nonumber \\
\theta &=& \bar \theta +a \left[ 3\frac{\omega^3}{\omega_0}(1-2u)\sin \alpha \cos \alpha \right]\frac{\tau \cos \eta}{\sin \bar \theta}\,,  \nonumber \\
\phi &=& \bar \phi + a \left[2\omega \omega_0 -3\frac{\omega^3}{\omega_0}(1-2u)\frac{\cos^2\alpha}{\sin^2 \bar \theta}  \right]\tau\,.
\end{eqnarray}
For this orbit, the constants of the motion can be calculated from Eqs.~\eqref{K24}--\eqref{K27}  and the results are
\begin{eqnarray}\label{J9}
E&=& \frac{A_0^2}{N_0}-a \frac{\omega_0\,u}{N_0^3}\,\cos \alpha\,,  \nonumber\\
\Phi&=& r_0^2\,\omega\, \cos \alpha - 3\,a \frac{u\,A_0^2}{N_0^3}\,\cos^2 \alpha\,,  \nonumber\\
K&=& r_0^4\, \omega^2 - 2\,a \frac{r_0^2\,\omega_0\,A_0^2}{N_0^4}\,\cos \alpha\,,
\end{eqnarray}
where $K>0$ for an orbit around the Earth by Eq.~\eqref{K2bB}.

Equations~\eqref{J8} and~\eqref{J9} are derived in Appendix A.

\section{Tetrad Frame for the Tilted Spherical Orbit}

The spherical orbit~\eqref{J8} represents the geodesic world line of the test mass $m$  in Kerr spacetime linearized in $a$. Therefore, we specialize the general results of Marck for this specific orbit under consideration here in order to find the spatial frame for this orbit. Integrating Eq.~\eqref{K31}, we obtain
\beq\label{T1}
\Psi =\omega_0 \tau +\mathcal{O}(a^2)\,,
\eeq
where we have set the constant  of integration equal to zero for the sake of consistency with the Schwarzschild limit. Next, we must now work out $\tilde \lambda_{\hat 1} $, $\tilde \lambda_{\hat 2} $ and $\tilde \lambda_{\hat 3} $ for this spherical orbit. It follows from Eqs.~\eqref{K33}--\eqref{K37} that the coframe is given by
\begin{eqnarray}\label{T2}
\tilde \lambda_{\hat 1}&=&\frac{1}{A_0}\,dr-\Big(\frac{a\,A_0}{r_0\omega_0}\sin\alpha \cos\alpha \frac{\sin \eta}{\sin \bar \theta}\Big)\,d\theta -\Big(\frac{a\,A_0}{r_0 \omega_0}\sin^2\alpha \sin \eta \cos\eta\Big)\,d\phi\,,  \nonumber\\
\tilde \lambda_{\hat 2}&=&\lambda_{\hat 2}= -\frac{a}{r_0}\sin\alpha \cos\eta\,dt + \frac{a}{\omega_0 r_0^2}\sin \alpha \sin \eta\,dr + \Big(\frac{r_0\cos\alpha}{\sin \bar \theta}-\frac{a}{r_0\omega_0}Y_2\Big)\,d\theta + \Big(r_0 \sin \alpha \cos \eta +\frac{a}{r_0\omega_0}Y_3\Big)\, d\phi\,,\nonumber\\
\tilde \lambda_{\hat 3}&=&-r_0\omega_0Z_0\,dt + Z_2\, d\theta + Z_3\, d\phi\,, 
\end{eqnarray}
where 
\begin{eqnarray}\label{T3}
Y_2 &=& \frac{A_0^2\,\sin^2\alpha \cos\eta}{\sin \bar \theta} \left[\cos \eta +\left( \frac{3u}{N_0^2} \right)  \omega \tau \frac{\cos^2 \alpha \sin \eta}{\sin^2 \bar \theta}  \right]\,,\nonumber\\
Y_3 &=& \left(\frac{A_0}{N_0}\right)^2 \sin \alpha \cos\alpha [(1-3u)\cos\eta +3u\, \omega \tau \sin \eta]\,,\nonumber\\
Z_0 &=& \frac{A_0}{N_0}\left(1-\frac{a\omega_0 \cos\alpha}{N_0^2}\right)\,,\nonumber\\
Z_2 &=& -r_0 \frac{A_0}{N_0}\frac{\sin \alpha}{\sin \bar \theta}\left\{
\cos \eta + \left( \frac{A_0}{N_0} \right)^2 a\omega_0 \cos \alpha \left[ \frac{1-6u +6u^2}{u(1-2u)}\cos \eta 
+3\,\omega \tau \frac{\sin \eta \cos^2\alpha}{\sin^2\bar \theta}\right]
\right\}\,,
\nonumber\\
Z_3 &=& r_0 \frac{A_0}{N_0} \left[\cos\alpha -a\omega_0 \left(\frac{3u}{1-3u}\cos^2\alpha +\frac{1-3u}{u}\sin^2 \alpha \sin^2\eta \right)  \right]\,.
\end{eqnarray}

Let us now compute $\tilde \lambda^\mu{}_{\hat 1}$, $\tilde \lambda^\mu{}_{\hat 2}$ and $\tilde \lambda^\mu{}_{\hat 3}$.
Along the orbit, $\tilde \lambda^\mu{}_{\hat i}=g^{\mu\nu}\tilde\lambda_{\nu \hat i}$, where the inverse metric to linear order in $a$ is given by 
\begin{equation}\label{T4}
g^{00}=-\frac{1}{A^2}\,,\quad  g^{11}=A^2\,, \quad g^{22}=\frac{1}{r^2}\,,\quad
g^{33}=\frac{1}{r^2 \sin^2\theta}\,,\quad g^{03}=-2\frac{J}{r^3\,A^2}\,.
\end{equation} 

We find that the frame is given by
\begin{eqnarray}\label{T5}
\tilde \lambda_{\hat 1}&=&A_0\,\partial_r -\Big(\frac{a\,A_0}{r_0^3\omega_0}\frac{\sin\alpha \cos\alpha \sin \eta}{\sin \bar \theta}\Big)\,\partial_\theta -\Big(\frac{a\,A_0}{r_0^3 \omega_0}\frac{\sin^2\alpha \sin \eta \cos\eta}{\sin^2\bar \theta}\Big)\,\partial_\phi\,, \nonumber\\
\tilde \lambda_{\hat 2}&=& \frac{a}{r_0}\sin \alpha \cos \eta\,\partial_t + \frac{a\,A_0^2}{r_0^2 \omega_0}\sin\alpha\sin \eta\,\partial_r+\Big(\frac{\cos\alpha}{r_0 \sin \bar\theta}-\frac{a}{r_0^3\omega_0}Y_2\Big)\,\partial_\theta +\mathbb{B}_3\,\partial_\phi\,, \nonumber\\
\tilde \lambda_{\hat 3}&=& \mathbb{C}_0\,\partial_t + \frac{Z_2}{r_0^2}\,\partial_\theta +\mathbb{C}_3\,\partial_\phi\,,
\end{eqnarray}
where
\begin{eqnarray}\label{T6}
\mathbb{B}_3 &=&  \frac{\sin \alpha \cos \eta}{r_0 \sin \bar \theta}+\frac{a}{r_0^3 \omega_0\sin^2 \bar \theta}\left[Y_3-3 u\, \left(\frac{A_0}{N_0}\right)^2\,\frac{\sin^3\alpha \cos\alpha \sin \eta \cos^2\eta}{\sin^2 \bar \theta}\,\omega \tau \right]\,, \nonumber\\
\mathbb{C}_0 &=& \frac{r_0\omega_0}{A_0\,N_0}\left[1-3 a\omega_0 \cos \alpha \left(\frac{A_0}{N_0}\right)^2 \right]\,,\nonumber\\
\mathbb{C}_3 &=& \frac{1}{r_0^2 \sin^2 \bar \theta}\left( 1-2a\Theta \frac{\cos \bar \theta}{\sin \bar \theta} \right)Z_3 +\frac{2ar_0 \omega_0^3}{A_0^2}Z_0\,.
\end{eqnarray}
As before, with an abuse of notation, we have denoted frame vectors and coframe 1-forms  using the same symbol, namely, $\tilde \lambda_{\hat \alpha}$.
Moreover, we recall that  $\Theta$ is given in Eq.~\eqref{J6}, so that $a \Theta =\theta -\bar \theta$ and we have to first order in the Kerr rotation parameter $a$
\beq\label{T7}
\sin\theta = \sin \bar \theta +a \Theta \cos \bar \theta\,.
\eeq

To first order in $\alpha$,
\beq\label{T8}
\sin \alpha \approx \alpha \,,\qquad \cos \alpha \approx 1\,,\qquad \sin \bar \theta \approx 1
\eeq
and hence the spatial frame~\eqref{T5} simplifies such that one recovers previous results given in Ref.~\cite{MA}, which were obtained by a different method based on directly integrating in this case the equations in display~\eqref{I3}. In this connection, it is important to notice that three typographical errors occur in Ref.~\cite{MA} that must be corrected: In Eq.~(22) of Ref.~\cite{MA}, $r_0$ in the denominator of the last term must be replaced by the speed of light $c$. Furthermore, in Eqs.~(11) and (13) of Ref.~\cite{MA}, the temporal components must be divided by $c$.

\section{Measured Curvature Components for the Spherical Orbit}

Let us first express the projection of the Weyl tensor on the tetrad frame $\tilde \lambda_{\hat \alpha}$ associated with the spherical orbit under consideration here. We find that 
\beq\label{W1}
\tilde{\mathcal{E}}= \omega_0^2\, \left[
\begin{array}{ccc}
1-3\,\Gamma^2 & \Xi & 0\cr
\Xi & -2+3\,\Gamma^2 & 0\cr
0 & 0 & 1\cr
\end{array}
\right]\,,\qquad
\tilde{\mathcal{H}}= \omega_0^2\, \left[
\begin{array}{ccc}
(4-7u)\,\Upsilon & H & 0\cr
H & (4u-3)\,\Upsilon & 0\cr
0 & 0 & (3u-1)\,\Upsilon\cr
\end{array}
\right]\,.
\eeq
Here we have defined 
\beq\label{W2}
\Gamma := \sqrt{\frac{1-2u}{1-3u}}\,\left(1-\frac{a\,\omega_0\,\cos \alpha}{1-3u}\right)\,
\eeq
and 
\beq\label{W3}
H := -3\frac{\sqrt{u(1-2u)} }{1-3u}\,\left[ 1-\frac{1-u}{u(1-3u)}\,\,a\,\omega_0\,\cos \alpha  \right]\,,
\eeq
such that for $\alpha=0$, they reduce to $\gamma$ and $-3\gamma^2 \,\beta$, respectively, to linear order in $a/r_0$, see Eqs.~\eqref{K23A} and~\eqref{K23B}. Moreover, 
\beq\label{W4}
\Xi := -3\,\xi \,\frac{\sqrt{1-2u} (1+2u)}{1-3u}\,\sin \alpha \,\sin \eta\,,
\eeq
where $\xi$ is the dimensionless quantity $\rho$, defined in Eq.~\eqref{K2bA}, evaluated along the spherical orbit, namely, 
\beq\label{W5}
\xi :=\rho(r_0)=  \frac{J}{M\,r_0^2\,\omega_0}\,
\eeq
and 
\beq\label{W6}
\Upsilon := 3\,\xi\,\frac{\sqrt{u}}{1-3u}\,\sin \alpha \,\sin \eta\,.
\eeq
When $\alpha=0$, so that the inclination of the orbit vanishes, the spherical orbit turns into the circular equatorial orbit with $\Xi=\Upsilon=0$ and Eq.~\eqref{W1} reduces to Eq.~\eqref{K23f}.  

It is now straightforward to use Eqs.~\eqref{K23d}--\eqref{K23e} with $\Psi=\omega_0\,\tau$ from Eq.~\eqref{T1} to find the components of $\mathcal{E}$ and $\mathcal{H}$. The main results of this paper are thus
\begin{eqnarray}\label{W7}
\mathcal{E}_{\hat 1\hat 1}&=& \omega_0^2\,[1-3\,\Gamma^2\, \cos^2 (\omega_0\tau )]\,,\nonumber\\
\mathcal{E}_{\hat 1 \hat 2}&=&\omega_0^2\,\Xi\, \cos (\omega_0 \tau)\,,  \nonumber\\
\mathcal{E}_{\hat 1 \hat 3}&=& -3\, \omega_0^2\,\Gamma^2\,\sin (\omega_0 \tau)\, \cos (\omega_0 \tau)\,,  \nonumber\\
\mathcal{E}_{\hat 2 \hat 2}&=& \omega_0^2\,(3\,\Gamma^2-2)\,, \nonumber\\
\mathcal{E}_{\hat 2 \hat 3}&=& \omega_0^2\,\Xi\,\sin (\omega_0 \tau)\,,  \nonumber\\
\mathcal{E}_{\hat 3 \hat 3}&=&  \omega_0^2\,[1-3\,\Gamma^2\, \sin^2 (\omega_0\tau )]\,
\end{eqnarray}
and 
\begin{eqnarray}\label{W8}
\mathcal{H}_{\hat 1 \hat 1}&=& -\omega_0^2\,\Upsilon\,[1-3u-5(1-2u)\cos^2(\omega_0 \tau)]\,,  \nonumber\\
\mathcal{H}_{\hat 1 \hat 2}&=&\omega_0^2\,H\,\cos (\omega_0 \tau)\,, \nonumber\\
\mathcal{H}_{\hat 1\hat 3}&=& 5\,\omega_0^2\, (1-2u)\,\Upsilon\, \sin (\omega_0 \tau) \cos (\omega_0 \tau)\,, \nonumber\\
\mathcal{H}_{\hat 2 \hat 2}&=&\omega_0^2\,(4u-3)\,\Upsilon\,,   \nonumber\\
\mathcal{H}_{\hat 2\hat 3}&=& \omega_0^2\,H\,\sin (\omega_0 \tau)\,, \nonumber\\
\mathcal{H}_{\hat 3 \hat 3}&=&-\omega_0^2\,\Upsilon\,[1-3u-5(1-2u)\sin^2(\omega_0 \tau)]\,.
\end{eqnarray}
The tidal matrix~\eqref{W7} agrees with the results of Marck~\cite{Marck}, when Marck's tidal matrix is linearized  in angular momentum $J$ and specialized to the tilted spherical orbit under consideration in this paper. The expressions for $\mathcal{E}_{\hat 1 \hat 2}$ and 
$\mathcal{E}_{\hat 2 \hat 3}$ contain the Mashhoon--Theiss effect~\cite{MaTh}. 

It is important to note that for $a=0$, the spherical orbit reduces to the inclined circular orbit in the exterior Schwarzschild spacetime, which is depicted in Figure 2. Moreover, $\mathcal{E}$ and 
$\mathcal{H}$ for $a=0$ become independent of inclination $\alpha$ as a consequence of the spherical symmetry of the background gravitational field; indeed, they reduce to the corresponding quantities given in Eq.~\eqref{K21} when we set $a=0$. Let us recall here that in this case the components of $\mathcal{E}$ and  $\mathcal{H}$ are all periodic in proper time $\tau$. That is,  the relativistic tidal matrix varies with frequency $2\,\omega_0$ with respect to $\tau$, while the components of  
$\mathcal{H}$ are all off-diagonal and vary with frequency $\omega_0$ with respect to $\tau$. 

We recover from these results to first order in $a/r_0$ and with $\alpha=0$, the measured components of the curvature for the equatorial circular orbit in Kerr spacetime. Moreover, to first order in $\alpha$, Eq.~\eqref{W7} reduces to the results given in Ref.~\cite{MA}. It is important to note that even though there are secular terms proportional to $\tau$ in both the inclined spherical orbit~\eqref{J8} and the components of the spatial frame, cf.\ Eqs.~\eqref{T5} and~\eqref{T6}, the measured curvature components~\eqref{W7}--\eqref{W8} do not contain such terms. Indeed, except for $\mathcal{E}_{\hat 2 \hat 2}$ that is independent of time $\tau$, the rest involve  \emph{periodic}  variations with respect to $\tau$. In particular, the time-dependent components of the tidal matrix $(\mathcal{E}_{\hat a \hat b})$ contain Fourier terms with frequencies $2\,\omega_0$, $\omega+\omega_0$ and $\omega-\omega_0$ with respect to proper time $\tau$; similarly, the elements of $(\mathcal{H}_{\hat a \hat b})$ involve Fourier terms with frequencies  $\omega$, 
$\omega_0$, $2\,\omega_0+\omega$ and $2\,\omega_0-\omega$. Furthermore, while the relativistic tidal matrix naturally contains purely Newtonian terms as well as their general relativistic corrections, the gravitomagnetic components of the curvature, $(\mathcal{H}_{\hat a \hat b})$, all vanish as $1/c$, when we formally let $c\to \infty$.

It is interesting to consider the eigenvalues of the matrices $\mathcal{E}$ and $\mathcal{H}$. These  eigenvalues are invariant under similarity transformations of these matrices; therefore, we can calculate the eigenvalues just as well using $\tilde{\mathcal{E}}$ and $\tilde{\mathcal{H}}$.  It is straightforward to see from Eq.~\eqref{W1} that the eigenvalues of $\mathcal{E}$ are given to linear order in $a/r_0$ by
\beq\label{W9}
1-3\,\Gamma^2\,, \qquad -2+3\,\Gamma^2\,, \qquad 1\,,
\eeq
since  $\Xi^2$ is of second order in $a/r_0$ and can be neglected. 
It is remarkable that the eigenvalues of the tidal matrix $\mathcal{E}$ are independent of time. This circumstance is consistent with the notion that the Mashhoon--Theiss effect~\cite{MaTh} comes about as a consequence of the parallel propagation of the observer's frame. 

Similarly, the eigenvalues of 
$\mathcal{H}$ are given to linear order in $a/r_0$ by
\beq\label{W10}
(3u-1)\,\Upsilon\,, \qquad  \frac{1}{2}\,(1-3u)\,\Upsilon\pm H\,.
\eeq

\subsection{Relativistic Tidal Matrix $\mathcal{E}_{\hat a \hat b}$}

Inspection of the relativistic tidal matrix, Eq.~\eqref{W7}, reveals that, except  for $\mathcal{E}_{\hat 1 \hat 2}$ and $\mathcal{E}_{\hat 2 \hat 3}$ that are proportional to $\Xi$ and have dominant amplitudes proportional to $\omega_0^2\,\xi$, the other elements of the tidal matrix contain expected Newtonian and post-Newtonian terms.  In fact, \emph{the off-diagonal components of  $\mathcal{E}$ that are proportional to $\Xi$ represent the Mashhoon--Theiss effect~\cite{MaTh}}. Moreover, the other components contain expected Newtonian terms proportional to $\omega_0^2= GM/r_0^3$, a series of post-Newtonian gravitoelectric terms with dominant amplitude proportional to 
\beq\label{W10a}
\omega_0^2\,u= \frac{G^2M^2}{c^2\, r_0^4}\,
\eeq
and a series of post-Newtonian gravitomagnetic terms with dominant amplitude proportional to 
\beq\label{W10b}
\omega_0^3\,a\,\cos \alpha=\omega_0^2\,\frac{a\,\omega_0}{c^2}\,\cos \alpha = \frac{G^2M^2}{c^2\, r_0^4}\,\xi\,\cos \alpha\,.
\eeq
These Newtonian and dominant post-Newtonian amplitudes occur in the first detailed post-Newtonian (``pN") treatment of the tidal matrix presented  in Refs.~\cite{PMW, MPW}. However, as pointed out in Ref.~\cite{MPW}, if the frame that is employed is parallel transported, then in the post-Newonian approximation certain secular terms occur in the tidal matrix as well.  In Ref.~\cite{MPW}, a more complete treatment that would take the secular terms into account was deferred to a future publication. The promised secular terms, which are the short-term manifestations of the Mashhoon--Theiss  effect~\cite{MaTh}, are given in the following section. 

\subsection{Secular Terms in $\mathcal{E}_{\hat a \hat b}$}

According to general relativity, the response of a gradiometer in orbit about the Earth is the projection of the Earth's Riemann curvature tensor onto the parallel-transported tetrad frame of the gradiometer. We naturally assume that the spatial frame of the gradiometer is determined by three orthogonal gyroscopes, while its temporal axis is fixed by the gradiometer orbit. Let us first consider  the secular motion of the gyroscopes as they orbit the Earth. 

The motion of an ideal test gyroscope with spin $\mathbf{S}$ in orbit about a central rotating source of mass $M$ and angular momentum $\mathbf{J}$ has been thoroughly studied in the first post-Newtonian approximation~\cite{Ever} and the result can be expressed as
\beq\label{W10c}
\frac{d\mathbf{S}}{d\tau} = (\boldsymbol{\Omega}_{ge} + \boldsymbol{\Omega}_{gm})\times \mathbf{S}\,,
\eeq
where 
\beq\label{W10d}
\boldsymbol{\Omega}_{ge} = \frac{3}{2}\,\frac{GM}{c^2\,r^3}\, \boldsymbol{\ell}\,, \qquad  \boldsymbol{\Omega}_{gm} = \frac{G}{c^2\,r^5}[3\,(\mathbf{J} \cdot \mathbf{x})\,\mathbf{x}-\mathbf{J}\,r^2]\,,
\eeq
 $|\mathbf{x}|=r$ and $\boldsymbol{\ell} = \mathbf{x}\times \mathbf{v}$ is the specific angular momentum of the orbit of the gyroscope. Here, $\boldsymbol{\Omega}_{ge}$ is the (gravitoelectric) \emph{geodetic} precession frequency of the gyroscope, while $\boldsymbol{\Omega}_{gm}$ is its \emph{gravitomagnetic} precession frequency.  These precession frequencies have  been
directly measured via Gravity Probe B~\cite{Ever}, which involved four superconducting gyroscopes and a telescope that were launched on 20 April 2004 into a polar Earth orbit of radius 642 km aboard a drag-free satellite.

The post-Newtonian equation for the motion of the spin describes the slow precession of the spin vector, which is cumulative. That is, in contrast to the ``fast" orbital motion, the geodetic and gravitomagnetic motions of the spin vector are ``slow", with long periods proportional to $c^2$. We therefore expect that over a period of time $\tau$, corresponding to the duration of a satellite gradiometry experiment in an inclined spherical orbit of radius $r_0$, the motion of the spatial frame of the gradiometer would accumulate geodetic and gravitomagnetic precession angles of order 
\beq\label{W10e}
\frac{GM}{c^2\,r_0}\, \omega_0 \tau\,, \qquad  \frac{GJ}{c^2\,r_0^3}\, \tau\,,
\eeq
respectively. These results are consistent with Eq.~\eqref{K19} for the parallel propagation of the spatial frame along an equatorial circular geodesic orbit in the exterior Kerr spacetime. 

In practice, the projection of the Riemann tensor onto the tetrad frame of the gradiometer necessitates detailed calculations in which the symmetries of the Riemann tensor need to be taken into account; that is, $\mathcal{E}_{\hat a \hat b}=R_{\hat 0 \hat a \hat 0 \hat b}$, which is given by Eq.~\eqref{I4}. If, after such detailed calculations, any post-Newtonian secular terms survive in the final result, we would expect them to be of the type presented in display~\eqref{W10e} multiplied by $\omega_0^2$, since the dominant terms in the Riemann curvature tensor are Newtonian in origin and proportional to $\omega_0^2$. Thus possible secular terms in $\mathcal{E}_{\hat a \hat b}$ would be expected to be of order
\beq\label{W10f}
\frac{GM}{c^2\,r_0}\, \omega_0^3\, \tau= u\,\omega_0^3\,\tau\,, \qquad  \frac{GJ}{c^2\,r_0^3}\,\omega_0^2 \tau=\frac{a}{c^2}\,\omega_0^4\,\tau\,,
\eeq
respectively. 

It is an important consequence of general relativity (GR) that secular terms~\eqref{W10f} do \emph{not} survive for a gradiometer following the circular equatorial orbit in Kerr spacetime, see Eq.~\eqref{K22}. Moreover, GR predicts that even off the equatorial plane the (gravitoelectric) geodetic secular term does \emph{not} survive for a spherical orbit; however, the gravitomagnetic secular term remains in this case, see Eq.~\eqref{W34}. To arrive at these conclusions as well as the precise form of the Mashhoon--Theiss effect in this case requires detailed evaluation of $R_{\hat 0 \hat a \hat 0 \hat b}$ within the framework of GR, see Eqs.~\eqref{W32} and~\eqref{W33}.

\section{Mashhoon--Theiss Effect}

The Marck tetrad frame that we have used in this paper to illustrate the Mashhoon--Theiss (``M-T") effect is unique up to a \emph{constant} rotation of the spatial frame. Thus up to such a rotation, the M-T effect is given by the off-diagonal terms in the tidal matrix~\eqref{W7} given by
\beq\label{W11}
\mathcal{E}_{\hat 1 \hat 2}=\omega_0^2\,\Xi\, \cos (\omega_0 \tau)\,,\qquad \mathcal{E}_{\hat 2 \hat 3}=\omega_0^2\,\Xi\, \sin (\omega_0 \tau)\,,
\eeq
where $\Xi$ is defined by Eq.~\eqref{W4}.
\emph{The remarkable property of such basically relativistic terms is that their amplitudes are independent of the speed of light $c$.} Let us write these terms in the form 
\beq\label{W12}
\mathcal{E}_{\hat 1 \hat 2}=-\frac{3}{2}\,\omega_0^2\,\xi \,\sin \alpha \,\frac{\sqrt{1-2u} (1+2u)}{1-3u}\,\left\{\sin [(\omega-\omega_0)\tau +\eta_0]+ \sin [(\omega+\omega_0)\tau +\eta_0]\right\}\,
\eeq
and 
\beq\label{W13}
\mathcal{E}_{\hat 2 \hat 3}=-\frac{3}{2}\,\omega_0^2\,\xi \,\sin \alpha \,\frac{\sqrt{1-2u} (1+2u)}{1-3u}\,\left\{\cos [(\omega-\omega_0)\tau +\eta_0]-\cos [(\omega+\omega_0)\tau +\eta_0]\right\}\,.
\eeq
These expressions indicate the presence of a beat phenomenon involving frequencies $\omega$ and $\omega_0$ with a \emph{beat frequency}
\beq\label{W14}
\omega_F :=\omega-\omega_0\,.
\eeq
This Fokker frequency ($\approx 3\,u\,\omega_0/2$)  is the gravitational analog of the Thomas precession frequency; that is, the gravitoelectric Fokker precession is the geodetic precession frequency of an ideal gyroscope on a circular orbit in the field of a spherical mass $M$.

In this paper, we have used the approach developed by Marck~\cite{Marck} to illustrate the M-T effect. However, the original work of Mashhoon and Theiss~\cite{MaTh} was done before the work of Marck~\cite{Marck} and involved finding the solutions to the equations in display~\eqref{I3} using a linear perturbation treatment. In the extensive calculations that had to be performed, one could see explicitly that the new effect came about due to a small denominator phenomenon involving the beat frequency $\omega_F$ in the calculation of the frame that is parallel transported along the orbit. That is, the near commensurability of frequencies $\omega$ and $\omega_0$ led to a small divisor that is ultimately responsible for the M-T effect. The phenomenon of small divisors is well known in celestial mechanics, since Laplace  in 1785 elucidated the commensurability of the mean motions of Jupiter and Saturn about the Sun.  

In connection with the origin of the M-T effect, let us first note that it is not intrinsic to the Kerr field; that is, the curvature of the Kerr field does not exhibit such a phenomenon, see Appendix B. The measured components of the curvature tensor basically involve the curvature tensor projected onto the tetrad frame of the observer. The small denominator (``resonance") phenomenon involving $\omega$ and $\omega_0$ described above that underlies the M-T effect must therefore originate in the parallel-propagated frame of the tilted spherical orbit, as there is no trace of a beat phenomenon in the orbital equations. The off-diagonal terms~\eqref{W11} are essentially ``Newtonian" in the sense that they do not vanish as $c\to\infty$; therefore, they can be combined with the diagonal Newtonian part of the tidal matrix via a \emph{constant} rotation such that at any given time the 
amplitude of the M-T effect can be reduced to zero. However, the tidal term then exhibits a beat phenomenon; that is, it is oscillatory with increasing amplitude and achieves its full tidal amplitude over a time comparable to the Fokker period $2\pi/\omega_F$. To see how this can come about, let us consider a constant rotation of the spatial frame given by  
\begin{eqnarray}\label{W15}
\lambda'_{\hat{1}}&=& \lambda_{\hat{1}} + \epsilon \sin \eta_0\,  \lambda_{\hat{2}}\,,\nonumber\\\lambda'_{\hat{2}}&=& -\epsilon \sin \eta_0 \,  \lambda_{\hat{1}} + \lambda_{\hat{2}} -\epsilon \cos \eta_0 \,  \lambda_{\hat{3}} \,,\nonumber\\
\lambda'_{\hat{3}}&=& \epsilon \cos \eta_0 \,  \lambda_{\hat{2}} + \lambda_{\hat{3}}\,,
\end{eqnarray}
where $\epsilon$,   $0<|\epsilon|<1$, is proportional to $a=J/M$ and hence will be treated to linear order. Under such a rotation $\mathbb{R}$, 
\beq\label{W16}
\mathcal{E'} = \mathbb{R}\,\mathcal{E}\,\mathbb{R}^{\dagger}\,,
\eeq
where 
\beq\label{W17}
\mathbb{R}= \left[
\begin{array}{ccc}
1 &  \epsilon\, \sin \eta_0 & 0\cr
- \epsilon\, \sin \eta_0 & 1 &  -\epsilon\, \cos \eta_0\cr
0 &  \epsilon\, \cos \eta_0 & 1\cr
\end{array}
\right]\,.
\eeq

To linear order in $a$, the only terms in ${\mathcal E}'$ that are different from ${\mathcal E}$ are the following off-diagonal terms
\begin{eqnarray}
{\mathcal E}'_{\hat 1 \hat 2} &=& \omega_0^2 \left[\Xi \cos (\omega_0 \tau) +3\epsilon (\Gamma^2-1)\sin \eta_0 + 3\epsilon \Gamma^2  \sin (\omega_0 \tau +\eta_0)\cos(\omega_0 \tau)  \right]\,, \label{W18} \\
{\mathcal E}'_{\hat 2 \hat 3} &=& \omega_0^2 \left[\Xi \sin (\omega_0 \tau) +2\epsilon (3\Gamma^2-2)\cos \eta_0 - 3\epsilon \Gamma^2  \cos (\omega_0 \tau +\eta_0)\cos(\omega_0 \tau)  \right]\,. \label{W19}
\end{eqnarray}
Let us write $\omega=\omega_F +\omega_0$, where $\omega_F$ is the Fokker frequency and note that $\Xi = \Xi_0 \sin \eta$, where
\beq\label{W20}
\Xi_0 :=-3\xi \frac{\sqrt{1-2u}(1+2u)}{1-3u}\sin \alpha\,.
\eeq
Moreover,  $\eta=(\omega_F \tau +\eta_0) + \omega_0 \tau$, hence
\beq\label{W21}
\sin \eta = \sin (\omega_F \tau +\eta_0)\cos (\omega_0 \tau)+\cos (\omega_F \tau +\eta_0)\sin (\omega_0 \tau)\,.
\eeq
If we choose $\epsilon$ such that
\beq\label{W22}
\Xi_0=-3\,\epsilon\, \Gamma^2\,,\qquad \epsilon= \xi\, \frac{1+2u}{\sqrt{1-2u}}\sin \alpha\,;
\eeq
then, we find
\begin{eqnarray}
\mathcal {E}'_{\hat 1 \hat 2} &=& \omega_0^2\, \Xi_0 \left[\mathbb{S}_F \cos^2 (\omega_0 \tau)+\mathbb{C}_F \sin  (\omega_0 \tau) \cos (\omega_0 \tau)-\frac{u}{1-2u}\sin \eta_0 \right]\,, \label{W23}
\\
\mathcal {E}'_{\hat 2 \hat 3} &=& \omega_0^2\, \Xi_0 \left[\mathbb{S}_F\sin  (\omega_0 \tau) \cos (\omega_0 \tau) +\mathbb{C}_F\sin^2 (\omega_0 \tau) -\frac{1+2u}{1-2u}\cos \eta_0\right]\,, \label{W24}
\end{eqnarray}
where the amplitudes of the \lq\lq fast" variation, with twice the Keplerian frequency, are given by
\begin{eqnarray}
\label{W25}
\mathbb{S}_F  &=& \sin (\omega_F \tau+\eta_0)-\sin \eta_0=2 \sin \left(\frac12\, \omega_F\, \tau  \right)  \cos \left(\frac12 \omega_F \tau  +\eta_0\right)\,,  \\
\label{W26}
\mathbb{C}_F  &=& \cos (\omega_F \tau+\eta_0)-\cos \eta_0=-2 \sin \left(\frac12\, \omega_F\, \tau  \right)  \sin \left(\frac12 \omega_F \tau  +\eta_0\right)\,. 
\end{eqnarray}
It follows that Eqs.~\eqref{W23} and~\eqref{W24} can be written as
\begin{eqnarray}
\mathcal {E}'_{\hat 1 \hat 2} &=& \omega_0^2\, \Xi_0 \left[2\,\sin \left(\frac{1}{2}\,\omega_F\,\tau\right)\,\cos (\omega'\,\tau) \cos (\omega_0 \tau)-\frac{u}{1-2u}\sin \eta_0 \right]\,, \label{W27}
\\
\mathcal {E}'_{\hat 2 \hat 3} &=& \omega_0^2\, \Xi_0 \left[2\,\sin \left(\frac{1}{2}\,\omega_F\,\tau\right)\,\cos (\omega'\,\tau)\,\sin  (\omega_0 \tau)  -\frac{1+2u}{1-2u}\cos \eta_0\right]\,, \label{W28}
\end{eqnarray}
where 
\beq\label{W29}
\omega' :=\frac12 (\omega+\omega_0)\,.
\eeq
The amplitude of the M-T effect is proportional to 
\beq\label{W30}
\sin \left(\frac12\, \omega_F\, \tau  \right)\,,
\eeq
which vanishes at $\tau=0$ and becomes unity at half the Fokker period, i.e., at proper time $\tau =\pi/\omega_F$.  Indeed, for a near-Earth orbit, $\xi \approx 3 \times 10^{-2}$ and  the Fokker period $2\pi/\omega_F$ is about  $10^5$ years.

For $\omega_F \tau \ll 1$, we have
\beq\label{W31}
\sin \left(\frac12\, \omega_F\, \tau  \right)\approx \frac34 \frac{GM}{c^2 r_0}\,\omega_0 \tau\,,
\eeq
so that for $\tau \ll \omega_F^{-1}$, Eqs.~\eqref{W27} and~\eqref{W28} can be written as
\beq\label{W32}
{\mathcal E}'_{\hat 1 \hat 2}=-\frac92 \frac{\sqrt{1-2u}\,(1+2u)}{1-3u}\,\omega_0^2\,\sin \alpha \left(  \frac{GJ\tau}{c^2 r_0^3}\right) \cos (\omega_0 \tau)\cos (\omega' \tau)-\omega_0^2\, \Xi_0\, \frac{u}{1-2u}\sin \eta_0
\eeq
and
\beq\label{W33}
{\mathcal E}'_{\hat 2 \hat 3} =-\frac92 \frac{\sqrt{1-2u}\,(1+2u)}{1-3u}\, \omega_0^2\,\sin \alpha \left(  \frac{GJ\tau}{c^2 r_0^3}\right) \sin (\omega_0 \tau)\cos (\omega' \tau)-\omega_0^2\, \Xi_0\, \frac{1+2u}{1-2u}\cos \eta_0\,.
\eeq
These results clearly bring out the short-term secular nature of the M-T effect that would be useful in any gravity gradiometry experiment. It is interesting to compare the dominant amplitude of the short-term secular M-T effect in Eqs.~\eqref{W32} and~\eqref{W33}, namely,  $\sim \omega_0^2\,\sin\alpha \,GJ\tau/(c^2\,r_0^3)$, with the first post-Newtonian periodic gravitomagnetic amplitude given in Eq.~\eqref{W10b}. The ratio of these amplitudes is $\omega_0\,\tau\, \tan \alpha$,
so that for $\tau \ll \omega_F^{-1}$, the M-T effect increases linearly with time, which is important for the experimental detection of this gravitomagnetic effect.

\subsection{Physical Interpretation of the M-T Effect}

The M-T effect has been discussed by Anandan~\cite{Anan},  Gill {\it et al.}~\cite{Gill} as well as Blockley and Stedman~\cite{BS}.  More recently, the M-T effect has received attention in connection with future satellite gradiometry experiments~\cite{Xu:2015nbq,  Qiang:2016kix,  Qiang:2016rlu}. It is therefore useful to recapitulate here the main features of the M-T effect that have been demonstrated in this paper. 

For a gravity gradiometer on an inclined ``circular" orbit about a central slowly rotating mass, the M-T effect shows up in the gravitomagnetic part of the relativistic tidal matrix when the local spatial frame is \emph{parallel transported} along the orbit. The nonrotating frame is fixed up to a \emph{constant} rotation; therefore, the appearance of the M-T effect can be adjusted by a constant rotation of the local frame. 

The M-T effect involves a subtle beat phenomenon involving the Fokker frequency $\omega_F = \omega-\omega_0$, which corresponds to the geodetic precession frequency of an ideal gyro on a circular orbit about a spherical mass.  

Suppose that by a constant rotation we set the M-T effect equal to zero at $\tau=0$. The M-T effect then consists of periodic terms that appear in certain components of the relativistic tidal matrix with an amplitude proportional to
\beq\label{W33a}
 6\, \omega_0^2\,\xi\, \sin\alpha \,\sin \left(\frac12\, \omega_F\, \tau  \right)\,,
 \eeq
where $\xi = J/(Mr_0^2\omega_0)$ is independent of the speed of light $c$.

The Fokker period for a near-Earth orbit is $\sim 10^5$ years. In any gravity gradiometry experiment, $\tau \ll 2\pi/\omega_F$; therefore, over the short term, the M-T effect appears as a first post-Newtonian secular gravitomagnetic contribution to the relativistic tidal matrix with amplitude   
\beq\label{W34}
\frac{9}{2}\, \omega_0^2\, \sin\alpha \,\frac{GJ}{c^2r_0^3\omega_0}\, (\omega_0\, \tau)=\frac{9\, a}{2\,c^2} \, \omega_0^4\,\tau\,\sin \alpha\,,
\eeq
cf.~Eqs.~\eqref{W32} and~\eqref{W33}. The corresponding secular term in the spatial frame is consistent with the first post-Newtonian gravitomagnetic precession of the frame~\cite{MA, BS}. 

It is possible to identify and study the gravitomagnetic terms proportional to $\xi$ in the parallel-propagated spatial frame that are responsible for the appearance of the M-T effect in the relativistic tidal matrix---see, for instance, Ref.~\cite{MA}. Such terms are not, however, of interest experimentally, since the relevant long-term periodic motion of the parallel-propagated spatial frame along the tilted spherical orbit clearly goes beyond the short-term gravitomagnetic precession that has been verified by the GP-B experiment~\cite{Ever}. The whole long-term motion of the frame is a periodic gravitomagnetic  nodding and has been termed ``relativistic nutation"~\cite{MA, Mashhoon:2000zt}.

\subsection{Comparison of the M-T Effect with Calculations using the First Post-Newtonian Approximation}

As already mentioned, the calculations by Mashhoon and Theiss~\cite {MaTh, T1, MT1} that 
originally led to the M-T effect were rather long and cumbersome.  As Mashhoon and Theiss~\cite {MaTh, T1, MT1} worked in the post-Schwarzschild approximation, thus taking the mass  $M$ of the source into account to all orders but the angular momentum $J$ only to first order, the same long-period results could presumably be obtained from the summation of an appropriate post-Newtonian series. To circumvent the details of the calculations, but illustrate how the small divisor (``resonance") phenomenon could possibly produce the effect, Gill {\it et al.} developed a simple model of the M-T effect~\cite{Gill}. We only wish to illustrate here the essential shortcoming of this approach, which has been recently adopted by Xu and Paik~\cite{Xu:2015nbq} with erroneous results. 

To illustrate the approach adopted by Gill {\it et al.}~\cite{Gill}, consider the equation for  the parallel propagation of a component of the spatial frame to first post-Newtonian (1pN) order. Gill {\it et al.} propose to  integrate this equation exactly; indeed,  their ``explanation" of what they called ``the Mashhoon--Theiss ``anomaly"" is based on this \emph{exact solution} of an equation that is valid only at the 1pN level. But in writing the original 1pN equation, they neglected the 2pN, 3pN,\ldots, terms, so that the only physical content of the correct solution of this equation should remain within the 1pN approximation scheme. Their exact solution is not logically consistent; hence, they do not have a correct explanation of the long-period M-T effect.  Their contention that ``there is no ``new" relativistic (resonant) effect related to rotating masses" (Gill {\it et al.}~\cite{Gill}, penultimate sentence of their abstract)  is therefore erroneous. 

More recently, Xu and Paik~\cite{Xu:2015nbq} erroneously claimed the existence of secular gravitoelectric terms in the relativistic tidal matrix due to the geodetic precession of the frame, see Eq.~(22a) in Xu and Paik~\cite{Xu:2015nbq}. However,  it follows from Eq.~\eqref{W7} that 
there are no secular gravitoelectric contributions to the tidal matrix; in fact, for $a=0$, the relativistic tidal matrix is simply periodic in $\tau$ with frequency $2\,\omega_0$, see Eq.~\eqref{K22}.

\subsection{Detection of the M-T Effect}

The M-T effect is \emph{not} an anomaly; indeed, as demonstrated in this paper, it is a direct consequence of general relativity within the post-Schwarzschild approximation scheme. \emph{In any experiment involving the Earth's gravity gradients, for instance, the mass and angular momentum of the Earth will naturally contribute to the result of the experiment to all orders}. The M-T effect is based on the exterior Kerr spacetime linearized in angular momentum.  To go beyond the linear order in angular momentum is conceptually straightforward, but our preliminary considerations indicate that it would involve rather long and complicated calculations. Such an endeavor is beyond the scope of the present work. 

To compare the M-T effect  with observational results in gravity gradiometry, it is necessary to take advantage of the fact that the predicted result is unique up to an arbitrary rotation of Marck's tetrad frame. Once the initial directions of the gyros are chosen as in, say, Marck's frame, it is necessary to introduce possible errors in the orientation of the orthonormal frame characterized by dimensionless parameters $\epsilon_i$, for $i=1,2,3$, as illustrated in Eq.~\eqref{W17} for the simple situation considered above. The M-T effect should then show up in time as a secular modulation of certain periodic terms of the relativistic tidal matrix.

\section{Post-Schwarzschild Approximation}

Gravitation can be identified with the curvature of spacetime according to the  general theory of relativity. The weakness of the gravitational interaction therefore makes it possible in most situations to treat gravitation as a small perturbation on flat Minkowski spacetime. The Newtonian approximation emerges as the zeroth-order perturbation that is independent of the speed of light. The post-Newtonian (pN) corrections then provide an approximation scheme for the weak-field and slow-motion situations in which the prediction of relativistic gravitation can be compared with observations. The pN framework has been employed in a wide variety of problems and it is widely expected to be adequate for the theoretical description of the results of experiments for the foreseeable future. In certain special circumstances, however, other approximation schemes can be developed; this paper has been about one  such possibility, namely, the post-Schwarzschild approximation and the comparison of its results with the pN framework.

Gravitational phenomena in the exterior vacuum region of an almost spherically symmetric mass distribution can be described in a post-Schwarzschild approximation scheme, since in the absence of any deviation from spherical symmetry the exterior field can be uniquely described by the Schwarzschild spacetime. For instance, if the central mass is slowly rotating, the Thirring--Lense term can be treated as a first-order perturbation on the Schwarzschild background. Therefore, in the first post-Schwarzschild approximation the proper rotation (or oblateness) of the body is considered to first order whereas {the mass of the central body is taken into account to all orders}. Compared to the standard pN approximation, the nonlinear character of general relativity is more strongly reflected in the post-Schwarzschild scheme. Mashhoon and Theiss developed the post-Schwarzschild approximation for the investigation of the relative (i.e., tidal) acceleration of two bodies orbiting a rotating central mass~\cite{MaTh,Theiss,T1}. The results have been used for the analysis of the tidal influence of the Sun on the Earth--Moon system~\cite{Mash,MA,MT1,MT2, Mashhoon:2000zt}.

The results of the post-Schwarzschild approximation described thus far pertain to a first-order rotational perturbation of the background Schwarzschild field. The question naturally arises whether similar results hold for other deviations of the source from spherical symmetry.  In fact, most astronomical bodies are oblate. The effect of oblateness, treated as a first-order static deformation of the source, has been investigated by Theiss~\cite{Theiss,T2} for the case of two test particles moving on a circular geodesic orbit of small inclination $\alpha$ about a central oblate body of mass $M$. In these calculations, the Erez--Rosen metric~\cite{ErRo} linearized in the quadrupole moment $Q$ has been employed. Theiss's calculations show that the contribution of the quadrupole moment $Q$ of the central mass to the gravity gradient along the orbit contains a relativistic part with a leading amplitude of the form
\beq\label{D1}
6\, \alpha\, \frac{Q\,c^2}{M\,r_0^4}\sin \left(\frac12 \,\omega_F\, \tau  \right)\,,
\eeq
which is similar to the case of gravitomagnetism, cf. Eq.~\eqref{W33a},  and shows a temporal variation with a frequency comparable with the Fokker frequency. This new relativistic effect can also be explained by the occurrence of a small divisor which shows up in the solution of the parallel transport equations~\cite{Theiss,T2}.  For $\tau \ll 1/\omega_F$,  the above amplitude reduces to a Newtonian expression of order $\alpha GQ\omega_0 \tau/r_0^5$.  It should be mentioned that, as in the gravitomagnetic case,  the relativistic quadrupole contributions to the tidal acceleration strongly depend upon the choice of the local inertial frame of reference. In Ref.~\cite{T2}, this frame has been chosen so as to cancel the resonance-like terms in the tidal matrix at $\tau=0$. Further discussion of this effect is contained in Ref.~\cite{MT2}.

\section{DISCUSSION}

In relativistic gravity gradiometry, one measures the elements of the relativistic tidal matrix, which is theoretically obtained in general relativity via the projection of spacetime curvature tensor on the nonrotating tetrad frame of an observer. In a gravity gradiometry experiment on a space platform in orbit about the Earth, the mass $M_\oplus$, angular momentum $J_\oplus$, quadrupole moment $Q_\oplus$ and higher moments of the Earth are all expected to contribute to the result of the experiment. For geodesic orbits in the exterior Kerr spacetime, Marck has calculated the relativistic tidal matrix~\cite{Marck}. We employ Marck's results in this paper to linear order in angular momentum $J$ in order  to determine the relativistic tidal matrix for an observer following an inclined ``circular" geodesic orbit about a slowly rotating spherical mass $M$.  The result is then used to illustrate the Mashhoon--Theiss effect~\cite{MaTh}, which involves the long-period gravitomagnetic part of the relativistic tidal matrix as well as subtle cumulative effects
 that can be measured in principle via relativistic gravity  gradiometry. 

\appendix

\section{Derivation of the Tilted Spherical Orbit}

Let us substitute Eq.~\eqref{J4} in the first integrals of the geodesic equation, namely, Eqs.~\eqref{K24}--\eqref{K27}, keeping only terms that are at most linear in $a/r_0$. We want to find 
$T$, $\Theta$ and $F$ such that at $\tau=0$,  $T(0)=\Theta(0)=F(0)=0$. 

The substitution of $t=(\omega/\omega_0)\,\tau + a\,T$ in Eq.~\eqref{K24} simply results in the relation $T=C\,\tau$, where $C$ is a constant given by
\begin{equation}\label{A1}
C=\frac{1}{A_0^2}\,(E_1-2u\,\omega\,\cos \alpha)\,,
\end{equation}
where $E=E_0+a\,E_1$ and we recall that $E_0=A_0^2/N_0$. Next, Eq.~\eqref{K25} with $r=r_0$ implies that 
\begin{equation}\label{A2}
K_1=\frac{2\,r_0^2}{N_0}\,(E_1-\omega\,\cos \alpha)\,,
\end{equation}
where $K=K_0+a\,K_1$ and we recall that $K_0=r_0^4\,\omega^2$. 
 
To linear order in $a$, Eq.~\eqref{K26} reduces to
\begin{equation}\label{A3}
\sin^2\theta \, \left(\frac{d\theta}{d\tau}\right)^2=(\,\omega^2+a\,r_0^{-4}\tilde{K}\,)\,\sin^2\theta - \omega^2 \cos^2\alpha -2\,\frac{a\,\omega}{r_0^2}\cos \alpha\, \Phi_1\,,
\end{equation}
where 
\begin{equation}\label{A4}
\tilde{K}=K_1 +2\, \frac{\,r_0^2}{N_0}\,A_0^2\,\omega\,\cos \alpha\,,
\end{equation}
$\Phi=\Phi_0+a\,\Phi_1$ and $\Phi_0=r_0^2\,\omega \,\cos \alpha$.  In Eq.~\eqref{A3}, we substitute  $\theta=\bar\theta +a\, \Theta(\tau)$.                   It proves useful to introduce 
$\sin\bar{\theta}~ \Theta := \mathbb{D}$; then, after some algebra, we get 
\begin{equation}\label{A5}
r_0^2\,\omega \sin \alpha \cos \eta\,\left(\frac{d\mathbb{D}}{d\tau} + \omega \tan \eta \, \mathbb{D}\right) = \omega \cos \alpha \,\Phi_1 -\frac{1}{2\,r_0^2}\,\tilde{K}\,\sin^2{\bar{\theta}}\,.
\end{equation}
This equation has the solution 
\begin{equation}\label{A6}
\mathbb{D}=\mathbb{D}_0\,\tau\,\cos\eta\,,
\end{equation}
where 
\begin{equation}\label{A7}
\tilde{K}\,\cos\alpha= 2r_0^3\,\omega\, \Phi_1\,, \qquad \tilde{K}\,\sin\alpha= -2r_0^4\,\omega \,\mathbb{D}_0\,.
\end{equation}
Finally, we substitute $\phi=\bar\phi +a F$ in Eq.~\eqref{K27}. The introduction of 
\begin{equation}\label{A8}
F=2\,\omega\,\omega_0\,\tau +\frac{\mathbb{L}}{\sin^2{\bar{\theta}}}
\end{equation}
leads to much simplification. We find that 
\begin{equation}\label{A9}
\mathbb{L}= \frac{1}{r_0^2} \Phi_1\,\tau\,.
\end{equation}

Putting all these results together, we see that we have the solution for the orbit, but of the three unknown constants $E_1$, $\Phi_1$ and $K_1$, only two are determined. To find the last remaining relation, we must go back to the geodesic equation $D \lambda^\mu{}_{\hat 0}/d\tau=0$, namely, 
\begin{equation}\label{A10}
\frac{d^2x^\mu}{d\tau^2} + \Gamma^\mu_{\alpha \beta}\,\frac{dx^\alpha}{d\tau}\,\frac{dx^\beta}{d\tau}=0 \,,
\end{equation}
where $x^\mu = ( t, r, \theta, \phi)$ in Boyer--Lindquist coordinates. For $r=r_0$, the radial component of the geodesic equation reduces to 
\begin{equation}\label{A11}
\left(\frac{d\theta}{d\tau}\right)^2+ \sin^2\theta \, \left(\frac{d\phi}{d\tau}\right)^2= \omega^2+2\,a\,\omega\,\omega_0\,(C-\omega\,\cos\alpha)\,.
\end{equation}
Substituting our solution in this equation, we find, after much algebra, that 
\begin{equation}\label{A12}
\mathbb{D}_0= -\omega_0 \,(C-3\,\omega\,\cos\alpha)\,\sin\alpha\,.
\end{equation}
With this additional equation, the spherical orbit is fully determined  and we recover Eqs.~\eqref{J8} and~\eqref{J9}.

\section{Curvature of Kerr Spacetime as Measured by Static Observers}

It turns out that the curvature of the Kerr field can be represented by 
\beq\label{B1}
\mathcal{E}= \mathbb{E}\, \left[
\begin{array}{ccc}
-2 & 0 & 0\cr
0 & 1 & 0\cr
0 & 0 & 1\cr
\end{array}
\right]\,,\qquad
\mathcal{H}= \mathbb{H}\, \left[
\begin{array}{ccc}
-2 & 0 & 0\cr
0 & 1 & 0\cr
0 & 0 & 1\cr
\end{array}
\right]\,,
\eeq
with respect to the canonical Petrov tetrad of the Kerr field~\cite{cart}. Here,
\beq\label{B2}
\mathbb{E}+i\,\mathbb{H} = \frac{M}{(r+i\,a\,\cos \theta)^3}\,.
\eeq
The Kerr field is of  type D in the Petrov classification and this accounts for the ``parallelism"
between the gravitoelectric and gravitomagnetic components of its curvature in Eq.~\eqref{B1}.  To elucidate this feature of the Kerr spacetime further, it is interesting to study the curvature of the Kerr field as measured by the \emph{static}  family of accelerated observers  with adapted frame
\beq\label{B3}
e_{\hat 0}=\frac{1}{\sqrt{-g_{tt}}}\partial_t\,,  \qquad e_{\hat 1}=\frac{1}{\sqrt{g_{rr}}} \partial_r\,, \qquad  e_{\hat 2}=\frac{1}{\sqrt{g_{\theta\theta}}} \partial_\theta\,,
\qquad e_{\hat 3}=\frac{1}{\sqrt{g_{\phi\phi}-\frac{g_{t\phi}^2}{g_{tt}}}}\left(-\frac{g_{t\phi}}{g_{tt}} \partial_t+\partial_\phi\right)\,,
\eeq
where the tetrad axes are primarily along the  Boyer--Lindquist coordinate directions. We recall that  Kerr metric~\eqref{K1} is given by $-ds^2=g_{\mu \nu} dx^\mu dx^\nu$; that is, 
\beq\label{B3a}
-\frac{g_{t\phi}}{g_{tt}} = -2\,\frac{Jr}{\Sigma -2Mr}\, \sin^2\theta\,, \qquad  \sqrt{g_{\phi\phi}-\frac{g_{t\phi}^2}{g_{tt}}}= \sqrt{\frac{\Sigma \Delta}{\Sigma -2Mr}}\,\sin \theta\,, \qquad \sqrt{-g}= \Sigma\,\sin \theta\,.
\eeq
We assume that $\theta \ne 0, \pi$. Moreover, static observers only exist in the exterior Kerr spacetime outside the \emph{stationary limit surface} given by  
\beq\label{B3b}
\Sigma -2Mr = \Delta -a^2\,\sin^2 \theta = 0\,.
\eeq
With respect to these static observers, the nonvanishing components of the tidal matrix are given by 
\begin{eqnarray}\label{B4}
\mathcal {E}_{\hat 1 \hat 1}&=& -2 \mathbb{E} \,\frac{\Delta +\frac{1}{2}\,a^2 \sin^2 \theta}{\Delta -a^2\sin^2 \theta}\,,\nonumber\\
\mathcal {E}_{\hat 1 \hat 2}&=& -3 a\,\sin \theta\, \mathbb{H} \,\frac{\Delta^{1/2}}{\Delta-a^2\,\sin^2 \theta}\,,\nonumber\\
\mathcal {E}_{\hat 2 \hat 2}&=&  \mathbb{E} \,\frac{\Delta + 2\,a^2 \sin^2 \theta}{\Delta -a^2\sin^2 \theta}\,,\nonumber\\
{\mathcal E}_{\hat 3 \hat 3}&=& \mathbb{E}\,,
\end{eqnarray}
where
\beq\label{B5}
 \mathbb{E} = \frac{M r (r^2-3 a^2\cos^2 \theta)}{\Sigma^3}\,, \qquad 
 \mathbb{H}= -\frac{M a (3r^2- a^2\cos^2 \theta)\,\cos \theta }{\Sigma^3}\,.
\eeq
Moreover, the nonzero elements of the gravitomagnetic part of the Weyl curvature are given by
\begin{eqnarray}\label{B6}
\mathcal {H}_{\hat 1 \hat 1}&=& -2 \mathbb{H} \,\frac{\Delta +\frac{1}{2}\,a^2 \sin^2 \theta}{\Delta -a^2\sin^2 \theta}\,,\nonumber\\
\mathcal {H}_{\hat 1 \hat 2}&=& 3 a\,\sin \theta \,\mathbb{E} \,\frac{\Delta^{1/2}}{\Delta-a^2\,\sin^2 \theta}\,,\nonumber\\
\mathcal {H}_{\hat 2 \hat 2}&=&  \mathbb{H} \,\frac{\Delta + 2\,a^2 \sin^2 \theta}{\Delta -a^2\sin^2 \theta}\,,\nonumber\\
{\mathcal H}_{\hat 3 \hat 3}&=& \mathbb{H}\,.
\end{eqnarray}
It is interesting to note that for $r\gg M$ and $r\gg a$, the off-diagonal components have the asymptotic expressions
\beq\label{B7}
 \mathcal{E}_{\hat 1 \hat 2} \sim 9\,\frac{Ma^2\,\sin \theta\,\cos \theta}{r^5}\,
\eeq
and
\beq\label{B8}
\mathcal{H}_{\hat 1 \hat 2} \sim 3\,\frac{Ma\,\sin \theta}{r^{4}}\,.
\eeq

These same electric and magnetic components of the curvature tensor given by Eqs.~\eqref{B4} and~\eqref{B6} were presented by us in a rather different context in Appendix B of Ref.~\cite{Bini:2015xqa}. Comparing the results given here with those in Ref.~\cite{Bini:2015xqa}, we note that the sign of the expression for $ \mathcal{E}_{\hat 1 \hat 2}$  should be changed in our previous work; moreover, similar sign errors have occurred there in the magnetic components of curvature that must be corrected. 

\begin{acknowledgments}
D.B. thanks ICRANet for partial support. 
\end{acknowledgments}

\end{document}